# Structure and Properties of α-NaFeO$_2$-type Ternary Sodium Iridates

Kristen Baroudi [a], Cindi Yim [a], Hui Wu [b,c], Qingzhen Huang [b], John H. Roudebush [a], Eugenia Vavilova [d,e], Hans-Joachim Grafe [d], Vladislav Kataev [d], Bernd Buechner [d,f], Huiwen Ji [a], Changyang Kuo [g], Zhiwei Hu [g], Tun-Wen Pi [g], Chiwen Pao [h], Jyhfu Lee [h], Daria Mikhailova [g], Liu Hao Tjeng [g] and R. J. Cava [a]

[a] Department of Chemistry, Princeton University, Princeton, NJ 08544 USA

[b] NIST Center for Neutron Research Gaithersburg, MD 20899-6102, USA

[c] Department of Materials Science and Engineering, University of Maryland, College Park, MD 20742-2115, USA

[d] Leibniz Institute for Solid State and Materials Research IFW Dresden, Germany

[e] Zavoisky Physical Technical Institute, Russian Academy of Sciences, Kazan, Russia

[f] Institut für Festkörperphysik, Technische Universität Dresden, Dresden, Germany

[g] Max Planck Institute for Chemical Physics of Solids, Nöthnitzer Straße 40, 01187 Dresden, Germany

[h] National Synchrotron Radiation Research Center, Hsinchu 30076, Taiwan, R.O.C.

**Abstract**

The synthesis, structure, and elementary magnetic and electronic properties are reported for layered compounds of the type Na$_{3-x}$MIr$_2$O$_6$ and Na$_{3-x}$M$_2$IrO$_6$, where M is a transition metal from the 3$d$ series (M=Zn, Cu, Ni, Co, Fe and Mn). The rhombohedral structures, in space group R-3m, were determined by refinement of neutron and synchrotron powder diffraction data. No clear evidence for long range 2:1 or 1:2 honeycomb-like M/Ir ordering was found in the neutron

Corresponding author: kbaroudi@princeton.edu (Kristen Baroudi)

powder diffraction patterns except in the case of M = Zn, and thus in general the compounds are best designated as sodium deficient α-NaFeO$_2$-type phases with formulas Na$_{1-x}$M$_{1/3}$Ir$_{2/3}$O$_2$ or Na$_{1-x}$M$_{2/3}$Ir$_{1/3}$O$_2$. Synchrotron powder diffraction patterns indicate that several of the compounds likely have honeycomb in-plane metal-iridium ordering with disordered stacking of the layers. All the compounds are sodium deficient under our synthetic conditions and are black and insulating. Weiss constants derived from magnetic susceptibility measurements indicate that Na$_{0.62}$Mn$_{0.61}$Ir$_{0.39}$O$_2$, Na$_{0.80}$Fe$_{2/3}$Ir$_{1/3}$O$_2$, Na$_{0.92}$Ni$_{1/3}$Ir$_{2/3}$O$_2$, Na$_{0.86}$Cu$_{1/3}$Ir$_{2/3}$O$_2$, and Na$_{0.89}$Zn$_{1/3}$Ir$_{2/3}$O$_2$ display dominant antiferromagnetic interactions. For Na$_{0.90}$Co$_{1/3}$Ir$_{2/3}$O$_2$ the dominant magnetic interactions at low temperature are ferromagnetic while at high temperatures they are antiferromagnetic; there is also a change in the effective moment. Low temperature specific heat measurements (to 2 K) on Na$_{0.92}$Ni$_{1/3}$Ir$_{2/3}$O$_2$ indicate the presence of a broad magnetic ordering transition. X-ray absorption spectroscopy shows that iridium is at or close to the 4+ oxidation state in all compounds. $^{23}$Na nuclear magnetic resonance measurements comparing Na$_2$IrO$_3$ to Na$_{0.92}$Ni$_{1/3}$Ir$_{2/3}$O$_2$ and Na$_{0.89}$Zn$_{1/3}$Ir$_{2/3}$O$_2$ provide strong indications that the electron spins are short-range ordered in the latter two materials. Na$_{0.62}$Mn$_{0.61}$Ir$_{0.39}$O$_2$, Na$_{0.80}$Fe$_{2/3}$Ir$_{1/3}$O$_2$, Na$_{0.90}$Co$_{1/3}$Ir$_{2/3}$O$_2$, Na$_{0.92}$Ni$_{1/3}$Ir$_{2/3}$O$_2$, Na$_{0.86}$Cu$_{1/3}$Ir$_{2/3}$O$_2$ and Na$_{0.89}$Zn$_{1/3}$Ir$_{2/3}$O$_2$ are spin glasses.

CSD-numbers:

Na$_{0.62}$Mn$_{0.61}$Ir$_{0.39}$O$_2$: 426657   Na$_{0.80}$Fe$_{2/3}$Ir$_{1/3}$O$_2$: 426659 Na$_{0.90}$Co$_{1/3}$Ir$_{2/3}$O$_2$: 426658

Na$_{0.92}$Ni$_{1/3}$Ir$_{2/3}$O$_2$: 426656  Na$_{0.86}$Cu$_{1/3}$Ir$_{2/3}$O$_2$: 426655  Na$_{2.8}$ZnIr$_2$O$_6$: 426660

**Introduction**

Compounds based on iridium oxide are of significant current interest due to the unusual magnetic and electronic properties that they can exhibit. These properties arise from the comparable strengths of spin-orbit coupling and electron correlations for ions of iridium, a $5d$ transition metal. The competing interactions lead to exotic behavior such as a giant magnetoelectric effect and Mott Insulating properties in $Sr_2IrO_4$,[1] spin liquid behavior in $Na_4Ir_3O_8$,[2] and Mott insulating behavior in $Na_2IrO_3$.[3] One especially desirable structural feature expected to host unusual properties in iridates is a layered honeycomb lattice of $Ir^{4+}$.[4] In an α-$NaFeO_2$-based structure type, such compounds would have the ideal formulas $Na_3MIr_2O_6$. $Na_2IrO_3$[3] (i.e. $Na_3NaIr_2O_6$), for example, has a layered honeycomb lattice of spin ½ $Ir^{4+}$ ions, and may follow the Kitaev model for spin ½ ions on a honeycomb lattice, which can allow for a spin liquid state.[4-6] To the best of our knowledge, the only other compound currently known where a ternary element is found in the edge-sharing layer of Ir octahedra in an $Na_2IrO_3$-like structure is $Na_3Cd_2IrO_6$,[7] and the only property reported is its crystal structure. Here we report the synthesis of six iridium oxide compounds based on the honeycomb compound $Na_2IrO_3$, replacing the sodium atom in the middle of the iridium honeycomb layer by the $3d$ transition metals Mn, Fe, Co, Ni, Cu and Zn. These compounds were studied based on the hypothesis that their magnetic properties would be different from those observed for $Na_2IrO_3$, and would shed light on the magnetic character of Ir in a honeycomb lattice geometry.

**Experimental**

Our exploration of the compositions of potential α-$NaFeO_2$-like compounds in $Na(M,Ir)O_2$ systems (M = a $3d$ transition metal) led to the observation that two types of compositions are stable. For the late transition metals Co, Ni, Cu and Zn, compounds of the nominal composition $NaM_{1/3}Ir_{2/3}O_2$ were found, while for the earlier transition metals Fe and Mn, compounds of nominal composition $NaM_{2/3}Ir_{1/3}O_2$ were found. The $NaM_{1/3}Ir_{2/3}O_2$ type compounds allow for the possibility of Ir honeycombs in the layers of edge-sharing $(M,Ir)O_6$ octahedra. Testing the phase assemblages for different compositions indicated that variations in M to Ir ratios of more than 5% near these nominal 1:2 and 2:1 compositions were not stable under our synthetic conditions. The compounds for final study were synthesized via solid state methods, in all cases from starting nominal compositions of $Na_3MIr_2O_6$ (M=Co, Ni, Cu and Zn) or $Na_3M_2IrO_6$ (M=Mn and Fe). Dry $Na_2CO_3$ powder, Ir metal powder and a transition metal oxide ($MnO_2$, $Fe_2O_3$, $Co_3O_4$, $NiO_2$, CuO, and ZnO respectively) were mixed in a 3.15:2:1 Na:M:Ir ratio for the M = Fe and Mn compounds, and a 3.15:1:2 Na:M:Ir ratio for the M = Co, Ni, Cu, and Zn compounds. These ratios gave single phase products. Based on slight variation in peak positions in X-ray diffraction we believe that the sodium content is slightly variable for each compound. The reactants were ground into a homogeneously mixed powder in an agate mortar and pestle and pressed into pellets. The pellets were placed in alumina crucibles with alumina lids and heated in air between 750 °C and 900 °C. They were reground and reheated until there was no sign of $IrO_2$ impurity by laboratory X-ray diffraction (Bruker D8 Focus or Rigaku Miniflex diffractometer). Typically, pellets were heated for 48 hours, ground, pressed into new pellets and heated for another 48 hours. $NaZn_{1/3}Pt_{2/3}O_2$, used as a nonmagnetic standard for the specific heat measurements, was synthesized by grinding together stoichiometric amounts of $Na_2CO_3$, ZnO and Pt metal powder. The ground powder was pressed into a pellet, heated at 800 °C for 48 hours and then reground and heated two subsequent times at 800°C for 48 hours. A similar procedure was employed to synthesize $Na_2IrO_3$ as a comparison compound for the NMR studies.

Single crystals of the compound $Na_{1-x}Cu_{1/3}Ir_{2/3}O_2$ were grown by heating a sintered pellet of $Na_{0.86}Cu_{1/3}Ir_{2/3}O_2$ powder. The pellet was heated at 750 °C for 12 hours and then heated at hotter temperatures in 50 °C increments for 12 hours each up to 1100 °C. The samples were heated longer at 1050 °C and 1100 °C allow the crystals to grow larger. Small, black crystals (up to approximately 0.1 mm long) grew on top of the pellet.

Single phase powder samples were studied by neutron powder diffraction (NPD) at the NIST Center for Neutron Research, spectrometer BT-1, and synchrotron X-ray diffraction studies were performed at the Advanced Photon Source at Argonne National Laboratory on beamline 11BM. NPD patterns were collected at 4 K and 300 K with wavelength 1.5403 Å; a Cu(311) monochromator with a 90° take off angle and in-pile collimation of 60 minutes of arc were used. Data were collected over the range of 3 − 168° 2-Theta with a step size of 0.05°. Synchrotron X-ray diffraction patterns were collected at 295 K with a wavelength of 0.413029 Å or 0.413157 Å and a 2θ range of 0.5 degrees to 49.996 degrees with steps of 0.001 degree. The 300 K NPD patterns were refined using the Rietveld method in the program Fullprof.[8] The final compound compositions, which were Na-deficient compared to the ideal 1:1:2 α-NaFeO$_2$-type formula, were determined through the structural refinements. The single crystal X-ray diffraction study was conducted on a Bruker APEX II diffractometer using Mo Kα radiation.

Temperature dependent DC magnetization (M), AC magnetization (M'$_{AC}$), applied field ($\mu_0$H) dependent magnetization, heat capacity (C), and resistivity (ρ(T)) measurements were performed on a Quantum Design Physical Property Measurement System (PPMS). Zero field cooled (ZFC) DC magnetization measurements were taken from 2 to 300 K. Temperature dependent DC Magnetization (M) was also measured on a Quantum Design Magnetic Property Measurement System for the compounds $Na_{0.80}Fe_{2/3}Ir_{1/3}O_2$ and $Na_{0.90}Co_{1/3}Ir_{2/3}O_2$. Measurements of M vs. $\mu_0$H showed linear behavior for fields up to 1 Tesla in all cases. The susceptibility was defined as M/$\mu_0$H at $\mu_0$H = 1 Tesla for the temperature dependent magnetization measurements of $Na_{0.62}Mn_{0.61}Ir_{0.39}O_2$, $Na_{0.92}Ni_{1/3}Ir_{2/3}O_2$, $Na_{0.86}Cu_{1/3}Ir_{2/3}O_2$ and $Na_{0.89}Zn_{1/3}Ir_{2/3}O_2$. For the compounds $Na_{0.80}Fe_{2/3}Ir_{1/3}O_2$ and $Na_{0.90}Co_{1/3}Ir_{2/3}O_2$ the susceptibility was defined as M/$\mu_0$H at $\mu_0$H = 0.1 Tesla. M'$_{AC}$ measurements were taken at an applied DC field of 2 Oe and an AC field of 3.5 Oe at frequencies of 100, 1000 and 10000 Hz. Heat capacity was measured on a sintered pellet of $Na_{0.92}Ni_{1/3}Ir_{2/3}O_2$ in zero magnetic field from 100 to 2 K. A sintered pellet of $NaZn_{1/3}Pt_{2/3}O_2$ was also measured to use as a nonmagnetic analog. Samples were prepared for ρ(T) measurements by sintering a pellet of each compound and cutting it into a rough brick shape. Platinum wire leads were attached to the pellets using silver paint. ρ(T) was measured from 350 K down to the temperature where the sample resistance became too large (~$10^5$ ohms) for the measurement apparatus employed. X-ray absorption spectra (XAS) at the 3$d$ transition metal $L_{2,3}$ edges and the Ir-$L_3$ edge were recorded at the 08B and 17C beamlines, respectively, of the National Synchrotron Radiation Research Center (NSRRC) in Taiwan. [23]Na nuclear magnetic resonance (NMR) measurements on samples of $Na_{0.92}Ni_{1/3}Ir_{2/3}O_2$, $Na_{0.89}Zn_{1/3}Ir_{2/3}O_2$ and on the reference sample of $Na_2IrO_3$ were performed on a Tecmag pulse solid-state NMR spectrometer in a field of 7T in the temperature range 1.6 – 80 K. The NMR spectra were obtained by measuring the intensity of the Hahn echo versus magnetic field. The $T_1$ relaxation time was measured with the stimulated echo pulse sequence. Due to the occurrence of hydration in some cases, samples were kept in desiccators or under vacuum for storage.

**Results and Discussion**

*Structure*

Na$_2$IrO$_3$, the basis for the current materials, has the structural formula Na$_3$NaIr$_2$O$_6$, denoting the fact that it consists of layers of formula NaIr$_2$O$_6$ made from edge-sharing metal-oxygen octahedra, with the IrO$_6$ octahedra in a honeycomb array, separated by layers of Na ions. Therefore, the structure refinements for the Na$_3$MIr$_2$O$_6$ and Na$_3$M$_2$IrO$_6$ compounds were originally performed with a monoclinic cell in space group C2/c (no. 15) or C2/m (no. 12), the space groups reported for Na$_2$IrO$_3$.[3, 9] With the exception of the Zn variant, however, there is no clear evidence from the NPD data for the long range ordering of the Ir ions into a honeycomb array in the current compounds. The neutron diffraction patterns were instead satisfactorily indexed with the higher symmetry α-NaFeO$_2$-like rhombohedral cell in the space group R-3m (No. 166). This cell and symmetry reflect that the average structure has disordered M and Ir ions. The disorder may come from either in-plane disorder of the metals or stacking fault disorder, which on average makes the ions appear to be randomly mixed. Rietveld refinements were performed on the 300 K neutron patterns using the program FullProf. The refined cell constants for each compound are given in Table 1 and Table 2. The rhombohedral structure displays a three layer repeat of edge sharing (Ir,M)O$_6$ octahedra alternating with layers of octahedrally coordinated Na ions. Each of the compounds was modeled with one type of Na position in the Na layers, one type of O position, and the Ir and M atoms disordered over a single site. The crystal structure is shown in figure 1a. A generalized strain formulation with Laue class -3m1 was used for each refinement.

The neutron diffraction patterns are well indexed by the R-3m cell. However, some of the synchrotron patterns for these compounds show a broad peak where one would expect to see peaks arising from the presence of distinct, ordered iridium and transition metal ion positions within the layers. The compounds with a 1:2 nominal ratio of M to Ir show the largest such peaks (in the Q = 1.25-1.5 Å$^{-1}$ range, see figures 2 and 3). Such compounds have the correct ratio of transition metal to iridium to form ordered Ir honeycombs with the M atom in the middle of the honeycomb, but it is unclear at this time whether such in-layer ordering, if present, is obscured by stacking faults[10] or whether the M/Ir ions in the layers are really only short range ordered for the current materials. The 4 K powder neutron diffraction patterns did not show the presence of any new peaks when compared to those at 300 K for any of the compounds, indicating the absence of long range magnetic ordering at 4 K in all cases.

During the refinements, the oxygen content was held fixed at full occupancy and the transition metal to iridium ratio was held constant at the ratio of the chemical composition of the initial reaction. This gave excellent fits of the models to the neutron diffraction data, except for the case of $Na_{0.62}Mn_{0.61}Ir_{0.39}O_2$. The Mn to Ir ratio was allowed to vary after the 2:1 nominal ratio resulted in a poor fit and the oxygen ion was allowed to move off the special 6c position onto the 18h position (with 1/3 occupancy), based on the analysis of a difference map that showed that the oxygen position was disordered. With this disordered oxygen position, the fit was improved. Given the structural complexities often observed in Mn-based oxides, this compound appears to be worthy of more detailed study.

The sodium occupancy was allowed to refine freely for each compound, as were isotropic thermal parameters for each atom. The neutron diffraction pattern of the sample of $Na_{0.90}Co_{1/3}Ir_{2/3}O_2$ has peaks with shoulders that were modeled by adding a second phase to the refinement with the same space group but a slightly shorter *a* axis and a slightly longer *c* axis. The thermal parameters, oxygen position and sodium occupancy were refined separately for the two phases. The compound $Na_{0.80}Fe_{2/3}Ir_{1/3}O_2$ displayed similar shoulders on peaks, but in this case the second phase had smaller *a* and *c* axes. The lower set of tick marks in the center and lower panel of figure 2 show the minor phases. Extra peaks not indexed by the R-3m space group or an ordered unit cell are present in the synchrotron diffraction patterns (but not the neutron diffraction patterns) for $Na_{0.62}Mn_{0.61}Ir_{0.39}O_2$ and $Na_{0.90}Co_{1/3}Ir_{2/3}O_2$ in the upper and lower panels of figure 2. These peaks are hypothesized to be from hydrates that formed when the compounds were left outside of a desiccator for a prolonged period.

Based on peaks that are not explained by the rhombohedral cell, a case can be made for long range iridium ordering into a honeycomb lattice within the Zn-Ir-O layer in $Na_{0.89}Zn_{1/3}Ir_{2/3}O_2$. The neutron diffraction pattern of $Na_{0.89}Zn_{1/3}Ir_{2/3}O_2$ was therefore indexed in the space group C2/m (No. 12), one of those reported for $Na_2IrO_3$. The unit cell transformation between the rhombohedral and monoclinic cells is $(a'b'c')_{monoclinic} = (2,1,0/0,-3,0/-2/3,-1/3,-1/3)$ $(a,b,c)_{rhombohedral}$. The fit to the data in space group C2/m is found in figure 4. Several of the weak peaks that are present in the compound that cannot be indexed in the rhombohedral cell are well indexed in the monoclinic cell. A model of the crystal structure is shown in figure 1; (b) shows the zinc-iridium layers separated by a layer of sodium atoms while (c) shows the *ab* plane and the ordered honeycomb of iridium. This compound seems to have stacking faults, which were not modeled quantitatively. Their effect on the average structure is to make the Zn and Ir appear to be disordered. The disorder arises from stacking, not from within the transition metal layer. The disordered Zn and Ir in the average structure in this compound was modeled by having iridium partially on the zinc site and the zinc partially on the iridium site while maintaining a 1:2 Zn:Ir ratio. This was accomplished by using linear restraints relations in FullProf. The resulting mixing was at the level of about 8 %. The lattice parameters, atomic positions and isotropic thermal parameters were allowed to refine. The oxygen was constrained to full occupancy but the sodium occupancy was refined for both positions. This refinement gave the stoichiometry $Na_{2.8}ZnIr_2O_6$, slightly different from the refined stoichiometry using space group R-3m. The fit of the neutron pattern with space group C2/m has the same overall quality as the fit with space group R-3m but accounts for weak peaks that are not indexed by the rhombohedral cell, which leads us to conclude that for the Zn compound, there is ordering of the Zn and Ir in the honeycomb layer and that the formula is best represented as $Na_{2.8}ZnIr_2O_6$. To facilitate the comparison to the other analogs presented here, the Zn variant will be presented as $Na_{0.89}Zn_{1/3}Ir_{2/3}O_2$ where data is presented. The parameters of the monoclinic cell are tabulated in table 3.

*Single Crystal Diffraction*

One of the compounds in the study that was refined as having an average rhombohedral structure and therefore M/Ir disorder was studied by single crystal X-ray diffraction to test for the presence of stacking faults in large proportion, evidenced by the presence of streaks of diffracted intensity in the reciprocal lattice. In this regard, we chose the Cu case as a likely candidate because Cu in the divalent state, if present, would display a strong Jahn-Teller distortion of its $CuO_6$ octahedron, which would in turn lead to the tendency for Cu-Ir ordering in the triangular plane of edge-sharing octahedra. Figure 5 shows, however, that the crystal of $Na_{1-x}Cu_{1/3}Ir_{2/3}O_2$ has only bright reflections from the rhombohedral average structure, with no evidence of superlattice reflections or streaking between reflections along the $c^*$ direction. This indicates that the Cu and Ir atoms are indeed disordered over the same site in the plane and that the apparent disorder present in the average structure is not due to the presence of stacking faults. A quantitative structure refinement was not performed on the single crystal.

*Magnetism*

With the exception of $Na_{0.90}Co_{1/3}Ir_{2/3}O_2$, the temperature dependences of the inverse magnetic susceptibilities are linear at high temperatures for all compounds, indicating Curie Weiss behavior; Curie Weiss fits were performed for the high temperature range (Figures 6 and 7). A temperature independent contribution was subtracted from the susceptibility for each fit and is included in Table 4. Based on their negative Weiss temperatures, $Na_{0.62}Mn_{0.61}Ir_{0.39}O_2$, $Na_{0.80}Fe_{2/3}Ir_{1/3}O_2$, $Na_{0.92}Ni_{1/3}Ir_{2/3}O_2$, $Na_{0.86}Cu_{1/3}Ir_{2/3}O_2$, and $Na_{0.89}Zn_{1/3}Ir_{2/3}O_2$ show dominantly antiferromagnetic interactions. Due to the presence of more than one type of magnetic ion (i.e. Ir plus the transition elements) in most of the compounds, all of the values are reported per formula unit.

In the case of $Na_{0.90}Co_{1/3}Ir_{2/3}O_2$ there is a kink in the inverse magnetic susceptibility. The data above and below the kink can be fit with the Curie-Weiss law (lower panel of figure 6); the higher temperature data has a larger Curie constant and effective moment than the low temperature data. The Weiss temperature for the low temperature data is positive (showing dominantly ferromagnetic interactions) while it is negative for the higher temperature data. This indicates that a transition occurs near 200 K, perhaps a spin state transition in the Co ions.[11]

For each compound, a transition with a downturn in susceptibility is seen in the magnetic susceptibility at low temperatures (see insets Figures 6 and 7). We designate the temperature of the downturn in $\chi(T)$ as $T_f$; in all cases it is smaller than the Weiss temperature, indicating the presence of magnetic frustration. A balance of competing interactions has been observed in many iridates, and is also expected for triangular and honeycomb lattices and disordered magnetic materials in general. Curie constants, Weiss temperatures, magnetic transition temperatures, and effective moments for each compound are found in table 4. The ratio of the Weiss temperature to the transition temperature $\theta/T_f$ is also included as a measure of magnetic frustration; the larger the number, the more magnetically frustrated the compound.

The AC magnetization data for each compound can be seen in Figure 8. As the frequency of the AC field increases, the decrease in susceptibility that marks $T_f$ occurs at higher temperatures; this is a characteristic trait of spin glasses. The ratio of the change in transition temperature ($\Delta T_f$) to the transition temperature ($T_f$) times the log of the change in frequency ($\Delta \log \omega$) can be used to parameterize this dependence; the values of $\Delta T_f/T_f\Delta\log\omega$ for each compound are found in table 4. For the compound $Na_{0.92}Ni_{1/3}Ir_{2/3}O_2$ there are two peaks in the AC magnetization, one at 22.1 K that is frequency dependent and one at 6 K that is not frequency dependent. The frequency dependent peak at 22.1 K is only observed under low magnetic field: the peak at 22.1 K is absent under a 1 T field as seen in the inset in the upper panel of figure 7.

Rearranging the Curie-Weiss law to $C/(|\theta|*\chi) = T/|\theta| + 1$ provides a way to compare the normalized susceptibility of different compounds.[12, 13] Figure 9 shows a plot of $C/(|\theta|*\chi)$ as a function of $T/|\theta|$. For $\theta < 0$, data that follows the Curie-Weiss law should fall on a line with a y-intercept of 1 and a slope of 1. On this plot, increasingly stable antiferromagnetic fluctuations beyond the Curie-Weiss law on approaching $T_f$ are reflected by a positive deviation from Curie-Weiss behavior and increasingly stable ferromagnetic fluctuations on approaching $T_f$ are reflected by a negative deviation from Curie-Weiss behavior. Figure 9 shows the normalized inverse susceptibility of the five compounds with dominant antiferromagnetic interactions on one plot. ($Na_{0.90}Co_{1/3}Ir_{2/3}O_2$ has dominantly ferromagnetic interactions at low temperature and so is omitted.) The five plotted compounds have $\theta < 0$ and so are fit to a line with intercept 1. They each show negative deviations from Curie-Weiss behavior, indicating that ferromagnetic fluctuations increase as $T_f$ is approached from above. The Cu-containing material has the largest degree of ferromagnetic correlations while the Zn compound maintains nearly ideal Curie Weiss behavior down to the lowest temperatures relative to its Curie Weiss θ. $Na_{0.80}Fe_{2/3}Ir_{1/3}O_2$ has only positive deviations, indicating that it does not have ferromagnetic fluctuations.

Field dependent magnetization measurements taken at 2 K are shown for each compound in figure 10. $Na_{0.62}Mn_{0.61}Ir_{0.39}O_2$, $Na_{0.80}Fe_{2/3}Ir_{1/3}O_2$, $Na_{0.90}Co_{1/3}Ir_{2/3}O_2$, and $Na_{0.92}Ni_{1/3}Ir_{2/3}O_2$ display hysteresis in their field dependent magnetization measurements while $Na_{0.86}Cu_{1/3}Ir_{2/3}O_2$, and $Na_{0.89}Zn_{1/3}Ir_{2/3}O_2$ do not. $Na_{0.90}Co_{1/3}Ir_{2/3}O_2$ displays dominant ferromagnetic interactions at low temperature in the $\chi(T)$ behavior so hysteresis is not unexpected. However, $Na_{0.62}Mn_{0.61}Ir_{0.39}O_2$, $Na_{0.80}Fe_{2/3}Ir_{1/3}O_2$, and $Na_{0.92}Ni_{1/3}Ir_{2/3}O_2$ have dominant antiferromagnetic interactions according to the Curie-Weiss fits, so they would not be expected to have hysteresis in field dependent magnetization. The fact that hysteresis is seen and that the saturation magnetization is never reached supports spin glass behavior for these compounds. [14]

*Heat Capacity*

The heat capacity of $Na_{0.92}Ni_{1/3}Ir_{2/3}O_2$ was measured between 100 K and 2 K, which covers the temperature range where both peaks are seen in the AC magnetization measurement (Figure 8). The raw heat capacity data for $Na_{0.92}Ni_{1/3}Ir_{2/3}O_2$ does not show a peak in this temperature regime (Figure 11). The heat capacity of $NaZn_{1/3}Pt_{2/3}O_2$, a nonmagnetic compound with the same crystal structure,[15, 16] was used as a nonmagnetic analog to subtract the phonon contribution to the specific heat. The specific heats were normalized at a temperature higher than possible magnetic ordering. With this subtraction, a broad peak in the magnetic heat capacity with a maximum near 8 K remains, as seen in the inset of figure 11. Thus the feature seen in the AC magnetization measurements at 6 K appears to be associated with a magnetic ordering transition, while the higher temperature one reflects the spin glass transition; spin glasses do not show a peak in specific heat measurements. [14]

*Resistivity*

Resistivity measurements show that $\rho(T)$ for each of the compounds increases with decreasing temperature, signifying that they are insulating. Figure 12 plots log $\rho(T)$ for each compound as a function of inverse temperature. The resistivities are linear on such a plot, and all have similar slopes, indicating that they have similar transport energy gaps. An estimation of the transport activation energy from these plots was performed using $\rho = \rho_0 \exp(-E_g/2k_BT)$; the activation energies are tabulated in table 4. In contrast to $Na_2IrO_3$, the data for these compounds follows an Arrhenius law.[3] The activation energies range from 0.3 (Cu case) to 0.8 (Zn case) eV, indicating that the transition metal has a significant effect on the transport properties.

*X-ray Absorption Spectroscopy*

Figure 13 presents the M-$L_{2,3}$ x-ray absorption spectra of $Na_xM_{1/3}Ir_{2/3}O_2$ and $Na_xM_{2/3}Ir_{1/3}O_6$ together with reference compounds MO (M=Cu, Ni, Co, and Mn) and $Fe_2O_3$ for energy calibration. This absorption edge is generated by the excitation of a 2p core electron into the unoccupied 3d states, showing two intense white lines $L_3$ and $L_2$. They can be loosely assigned to $2p_{3/2}$ and $2p_{1/2}$ core hole final states, respectively. With increasing atomic number from Mn to Cu, the energy separation between $L_3$ and $L_2$ edges increases. The line shape of the spectrum depends strongly on the multiplet structure given by the intra-atomic Coulomb and exchange interactions of 3d transition metals, as well as the local crystal fields and the hybridization with the O 2p ligands. The XAS spectra at the 3*d* transition metal $L_{2,3}$ edges are highly sensitive to the valence state: an increase of the valence state of the metal ion by one causes a shift of the XAS $L_{2,3}$ spectra by one or more eV toward higher energies.[17] This shift is due to a final state effect in the x-ray absorption process.[17] The XAS spectra are also very sensitive to the spin state: different spin states result in different multiplet spectral structures.[18] From the energy position of the lines we can conclude that the Cu ion in $Na_{0.86}Cu_{1/3}Ir_{2/3}O_2$ and the Ni ion in $Na_{0.92}Ni_{1/3}Ir_{2/3}O_2$ are in the divalent state, while the Co ion in $Na_{0.90}Co_{1/3}Ir_{2/3}O_2$ and the Fe ion in $Na_{0.80}Fe_{2/3}Ir_{1/3}O_2$ are in the trivalent state. The splitting of the main peak at the Fe-$L_3$ edge of $Na_{0.80}Fe_{2/3}Ir_{1/3}O_2$ reflects the crystal field splitting 10 Dq in octahedral symmetry. The crystal field splitting in $Na_{0.80}Fe_{2/3}Ir_{1/3}O_2$ is 200 meV larger than that in reference compound $Fe_2O_3$ demonstrating a large crystal field interaction in $Na_{0.80}Fe_{2/3}Ir_{1/3}O_2$. Such a large crystal field interaction leads to a low spin (LS) state of the Co ion in $Na_{0.90}Co_{1/3}Ir_{2/3}O_2$, since the multiplet spectral structure at the Co-$L_{2,3}$ edges of $Na_{0.90}Co_{1/3}Ir_{2/3}O_2$ is the same as that of LS $Co^{3+}$ oxide $LiCoO_2$ found previously[19] (The weak feature at 777.8 eV indicates 6% of a $Co^{2+}$ phase in the sample). The main peak in the Mn-$L_3$ edge of $Na_{0.62}Mn_{0.61}Ir_{0.39}O_2$ is shifted by only 0.4 eV to lower energy relative to that $Mn^{4+}$ oxide $Sr_3Mn_2O_7$,[20] but 1.10 eV to higher energy with respect to that of $Mn^{3+}$ oxide $LaMnO_3$; indicating an average $Mn^{3.73+}$ valence state.

Figure 14 shows the Ir-$L_3$ XAS spectra of $Na_{0.86}Cu_{1/3}Ir_{2/3}O_2$ together with $IrCl_3$, $Li_2IrO_3$ and $Sr_2ScIrO_6$ as $Ir^{3+}$, $Ir^{4+}$ and $Ir^{5+}$ references. For 5$d$ transition-metal oxides, the broad 5$d$ band and large lifetime broadening overwhelm the weak multiplet interactions, so the XAS spectrum at the Ir-$L_{2,3}$ edge exhibits a single white line. Nevertheless, the energy position of the white line is still sensitive to the Ir valence state, as shown in Figure 14. We observe a higher energy shift of 1.65 eV from $Ir^{3+}$ in $IrCl_3$ to $Ir^{4+}$ in $Li_2IrO_3$ and further shift of 1.25 eV to $Ir^{5+}$ in $Sr_2ScIrO_6$. The white line at the Ir-$L_2$ edge of all the current compounds (shown only for $Na_{0.86}Cu_{1/3}Ir_{2/3}O_2$) lies at nearly the same energy of that of $Ir^{4+}$ in the compound $Li_2IrO_6$, indicating that the oxidation state of the iridium in all the current materials is close to 4+.

In the parent compound $Na_2IrO_3$ the sodium atom in the middle of the honeycomb has a 1+ oxidation state. The transition metals replacing sodium have oxidation states ranging from 2+ to 3.73+. This extra positive charge is accommodated by the sodium deficiency in these compounds. Based on a 4+ oxidation state for Ir, the electron configuration of Ir is expected to be $t_{2g}^5$, which is spin ½. This is important to note because the predictions made about iridium honeycombs apply to iridium ions with spin ½.[4, 5] The assignment of oxidation states based on the spectroscopy data is in general not clearly consistent with the formal valences that are implied from the nominal and refined formulas of the compounds. In the most straightforward of the current compounds, $Na_{2.8}ZnIr_2O_6$, a formal oxidation state of 3.6 is implied and yet seen in the spectroscopy as $Ir^{4+}$. The synthesis and spectroscopic investigation of an unambiguous oxide standard for $Ir^{3+}$ may help to resolve the discrepancies.

*Nuclear Magnetic Resonance*

Characteristic $^{23}$Na NMR spectra of $Na_2IrO_3$, $Na_{0.92}Ni_{1/3}Ir_{2/3}O_2$, and $Na_{0.89}Zn_{1/3}Ir_{2/3}O_2$ are shown in Figure 15 a, b and c. The spectra consist of a central inhomogeneously broadened line and, in certain cases, at high temperatures, a specific step-like structure is visible on the wings (Figure 15 a). This structure is due to a powder averaging of the quadrupole satellites arising due to the interaction of the $^{23}$Na spin $I = 3/2$ with the gradient of the crystal electrical field. (Note that the additional sharp peaks in the spectra are parasitic $^{63}$Cu and $^{27}$Al NMR signals from the probehead used as the markers for calibration of the magnetic field.)

In the reference compound $Na_2IrO_3$ a transition to a long-range ordered antiferromagnetic state at $T_N = 15$ K has been established by magnetization and specific heat measurements.[3] In particular, a sharp lambda-like anomaly in the specific heat at $T_N$ suggests a bulk nature of magnetic order.[3] Indeed, our NMR data on $Na_2IrO_3$ show a behavior rather typical for a 3D antiferromagnet. In the paramagnetic regime, the $^{23}$Na NMR spectrum reveals a temperature independent partially resolved quadrupole powder pattern with a practically constant width of the signal down to temperatures very close to $T_N$ (Figure 15 a, Figure 16 a). At the magnetic phase transition, the signal begins to broaden continuously down to the lowest measurement temperature of 1.6 K. Such inhomogeneous broadening is obviously related to the powder-averaged distribution of internal magnetic fields, static on the NMR timescale, acting on the $^{23}$Na nuclei. A sharp onset of magnetic order is further manifested in the $^{23}$Na nuclei relaxation rate $1/T_1$, which exhibits small and temperature independent values both above and below $T_N$ and a sharp peak at $T_N$ (Figure 16 b). Since the dominant relaxation channels of nuclear spins in a magnetic material are fluctuating dipoles and hyperfine fields arising from electron spins, the enhancement of $1/T_1$ takes place when the fluctuation rate of electron spins falls into the NMR frequency window in the MHz range, giving rise to a so-called Redfield peak.[21] In the paramagnetic state the fluctuations are too fast to be effective in relaxing the nuclear spins whereas in the ordered state the spins are static. Therefore the temperature evolution of the $^{23}$Na NMR spectral shape and of $1/T_1$ gives evidence for a rapid onset of the static magnetic order in $Na_2IrO_3$.

The situation is quite different in the $Na_{0.92}Ni_{1/3}Ir_{2/3}O_2$ material. Here the $^{23}$Na NMR spectrum broadens continuously starting from a high temperature of 80 K (Figure 15 b, Figure 16 a). Notably, the signal does not show an anomaly around 20 K where the frequency dependent peak has been observed in the AC susceptibility (Figure 8). This suggests that the spin-glass like behavior is suppressed in strong fields. However, the linewidth tends to saturate at $T_N \sim 6$ K where the static susceptibility has a maximum and the AC susceptibility shows a second frequency independent peak (Figure 8). The temperature dependence of $1/T_1$ is also in a stark contrast with that in $Na_2IrO_3$. The Redfield peak is much broader and is centered at $T_N \sim 6$ K in a close correspondence with the static magnetic data (Figure 16 b). Therefore our NMR data indicate the occurrence of a very broad temperature regime where the electron spins form short-range correlated regions. The spins exhibit a very slow dynamics and eventually freeze below $T_N \sim 6$ K in some kind of static and possibly disordered state. A rather similar temperature evolution of the $^{23}$Na NMR spectral shape and of the linewidth in the related Zn-substituted material $Na_{0.89}Zn_{1/3}Ir_{2/3}O_2$ (Figure 15 c, Figure 16 a) suggests a similar scenario of quasi-static short range spin correlations occurring in a broad temperature range.

**Conclusions**

The structures of six α-NaFeO$_2$-like compounds $Na_{0.62}Mn_{0.61}Ir_{0.39}O_2$, $Na_{0.80}Fe_{2/3}Ir_{1/3}O_2$, $Na_{0.90}Co_{1/3}Ir_{2/3}O_2$, $Na_{0.92}Ni_{1/3}Ir_{2/3}O_2$, $Na_{0.86}Cu_{1/3}Ir_{2/3}O_2$, and $Na_{0.89}Zn_{1/3}Ir_{2/3}O_2$ were modeled in the rhombohedral space group R-3m (no. 166). They are made up of three layers of edge sharing IrO$_6$ and MO$_6$ octahedra with layers of octahedrally coordinated Na atoms in between. The 1 : 2 metal atom ratios would allow for the formation of a honeycomb lattice but, the Ir and M atoms appear to be disordered over the same site, on average. There is evidence in the synchrotron diffraction patterns, however, indicating that some of the compounds display M/Ir ordering into a honeycomb lattice. For the compound $Na_{0.89}Zn_{1/3}Ir_{2/3}O_2$ it was shown that an ordered one layer monoclinic unit cell in space group C2/m (no. 12) is also a good fit to the diffraction data, indicating that this compound may have ordered iridium honeycombs. Magnetic measurements show that $Na_{0.62}Mn_{0.61}Ir_{0.39}O_2$, $Na_{0.80}Fe_{2/3}Ir_{1/3}O_2$, $Na_{0.92}Ni_{1/3}Ir_{2/3}O_2$, $Na_{0.86}Cu_{1/3}Ir_{2/3}O_2$, and $Na_{0.89}Zn_{1/3}Ir_{2/3}O_2$ have dominant antiferromagnetic interactions while $Na_{0.90}Co_{1/3}Ir_{2/3}O_2$ has dominant ferromagnetic interactions at low temperature and dominant antiferromagnetic interactions at high temperature. Magnetic ordering transitions occur well below the Weiss temperature for each of the compounds, pointing to magnetic frustration. XAS shows that even with the substitution of transition metals with oxidation state between 2+ and 3.73+ the oxidation state of iridium remains close to 4+, which suggests that the iridium is spin ½; however, quantitative interpretation of the effective moments observed does not allow for a simple single magnetic picture to emerge that describes the whole family. Based on the frequency dependence of the transition temperature in AC susceptibility measurements, hysteresis in field dependent magnetization measurements, lack of a peak in the heat capacity of $Na_{0.92}Ni_{1/3}Ir_{2/3}O_2$, and in the case of $Na_{0.92}Ni_{1/3}Ir_{2/3}O_2$ and $Na_{0.89}Zn_{1/3}Ir_{2/3}O_2$ NMR measurements, each of the compounds appear to be spin glasses. This behavior is distinct from the behavior seen in the parent compound $Na_2IrO_3$. However, a spin liquid state is not realized. The high sodium nonstoichiometry, the disordered crystal structure, and the apparently high Mn oxidation state in the compound with average formula $Na_{0.62}Mn_{0.61}Ir_{0.39}O_2$ suggest that there may be charge ordering and other complex magneto-structural phenomena present. The $Na_x(Mn,Ir)O_2$ system is clearly worthy of further study more detailed study, as are several of the other variants.

**Acknowledgements**

The work at Princeton was supported by the NSF Solid state chemistry program, grant NSF-DMR-1005438. The authors thank the 11-BM team at the Advanced Photon Source for their excellent synchrotron diffraction data. Use of the Advanced Photon Source at Argonne National Laboratory was supported by the U. S. Department of Energy, Office of Science, Office of Basic Energy Sciences, under Contract No. DE-AC02-06CH11357. EV acknowledges support of RFBR through Grant RFBR 14-02-01194. Jason Krizan is acknowledged for helpful discussions.

**Captions**

**Table 1:** shows the unit cell parameters, atomic positions, thermal parameters, and goodness of fit for the Rietveld refinements in space group R-3m of compounds $Na_{0.90}Co_{1/3}Ir_{2/3}O_2$ (major and minor phases), $Na_{0.92}Ni_{1/3}Ir_{2/3}O_2$, $Na_{0.86}Cu_{1/3}Ir_{2/3}O_2$, and $Na_{0.89}Zn_{1/3}Ir_{2/3}O_2$.

**Table 2:** shows the unit cell parameters, atomic positions, thermal parameters, and goodness of fit for the Rietveld refinements in space group R-3m of compounds $Na_{0.62}Mn_{0.61}Ir_{0.39}O_2$ and $Na_{0.80}Fe_{2/3}Ir_{1/3}O_2$ (major and minor phases).

**Table 3:** shows the unit cell parameters, atomic positions, thermal parameters, and goodness of fit for the Rietveld refinements in space group C2/m of the compound $Na_{2.8}ZnIr_2O_6$.

**Table 4:** for the compounds $Na_{0.62}Mn_{0.61}Ir_{0.39}O_2$, $Na_{0.80}Fe_{2/3}Ir_{1/3}O_2$, $Na_{0.92}Ni_{1/3}Ir_{2/3}O_2$, $Na_{0.90}Co_{1/3}Ir_{2/3}O_2$, $Na_{0.86}Cu_{1/3}Ir_{2/3}O_2$, and $Na_{0.89}Zn_{1/3}Ir_{2/3}O_2$ table 4 shows $\chi_0$, C, $\theta$ and $\mu_{eff}$ from the Curie Weiss fits, $T_f$, $\theta/T_f$ and $\Delta T_f/T_f\Delta\log\omega$ from the AC susceptibility measurements, band gap calculated from resistivity measurements, and the oxidation state determined by XAS. For the compound $Na_{0.90}Co_{1/3}Ir_{2/3}O_2$ the values from the high temperature Curie Weiss fit are also shown.

**Figure 1**: a) Representative average rhombohedral crystal structure of the $Na_{1-x}(M_yIr_{1-y})O_2$ phases. Red spheres represent oxygen, yellow spheres represent sodium, and purple spheres represent the transition metal / iridium. b) A view of the layers in the monoclinic unit cell. Red spheres represent oxygen, black spheres represent zinc, blue spheres represent iridium, and yellow spheres represent sodium. c) A view of the ab-plane of the monoclinic unit cell of the ordered $Na_{2.8}ZnIr_2O_6$ honeycomb compound. Red spheres represent oxygen, black spheres represent zinc, and blue spheres represent iridium.

**Figure 2**: For each panel: Observed intensity (red squares), calculated intensity (black line), difference calculation (blue line) and *hkl* reflections (green dashes). Upper panel: NPD pattern of $Na_{0.62}Mn_{0.61}Ir_{0.39}O_2$ collected at 300 K, indexed to the space group R-3m. Inset: a portion of a synchrotron diffraction pattern collected at 298 K blown up to show where there would be peaks in a unit cell with an ordered honeycomb lattice; the y axis is the measured intensity divided by the intensity of the largest peak to give a sense of scale. There is a small broad peak around Q = 1.5 $Å^{-1}$ that indicates the presence of a small amount of in-plane ordering. The sharp peaks marked with an asterisk are believed to be from a hydrate of this compound. Center panel: NPD pattern of $Na_{0.80}Fe_{2/3}Ir_{1/3}O_2$ collected at 300 K, indexed to the space group R-3m. Inset: a portion of a synchrotron diffraction pattern collected at 298 K blown up to show where there would be peaks in a unit cell with an ordered honeycomb; the y axis is the measured intensity divided by the intensity of the largest peak. No peaks are seen around Q = 1.5 $Å^{-1}$ that would indicate ordering. Lower panel: NPD pattern of $Na_{0.90}Co_{1/3}Ir_{2/3}O_2$ collected at 300 K, indexed to the space group R-3m. Inset: a portion of a synchrotron diffraction pattern collected at 298 K blown up to show where there would be peaks in a unit cell with an ordered honeycomb; the y axis is the measured intensity divided by the intensity of the largest peak. There is a broad peak around Q = 1.5 $Å^{-1}$ that indicates the presence of some in-plane ordering. The sharp peaks marked with an asterisk are believed to be from a hydrate.

**Figure 3**: For each panel: Observed intensity (red squares), calculated intensity (black line), difference calculation (blue line) and *hkl* reflections (green dashes). Upper panel: Neutron diffraction pattern of $Na_{0.92}Ni_{1/3}Ir_{2/3}O_2$ collected at 300 K, indexed to the space group R-3m. Inset: a portion of a synchrotron diffraction pattern collected at 298 K blown up to show where there would be peaks in a unit cell with an ordered honeycomb; the y axis is the measured intensity divided by the intensity of the largest peak. There is a broad peak around $Q = 1.5$ $Å^{-1}$ that indicates the presence of some in-plane ordering. Center panel: Neutron diffraction pattern of $Na_{0.86}Cu_{1/3}Ir_{2/3}O_2$ collected at 300 K with fit, indexed to the space group R-3m. Inset: a portion of a synchrotron diffraction pattern collected at 298 K blown up to show where there would be peaks in a unit cell with an ordered honeycomb; the y axis is the measured intensity divided by the intensity of the largest peak. There is a broad peak around $Q = 1.5$ $Å^{-1}$ that indicates the presence of some in-plane ordering. Lower panel: Neutron diffraction pattern of $Na_{0.89}Zn_{1/3}Ir_{2/3}O_2$ collected at 300 K, indexed to the space group R-3m. Inset: a portion of a synchrotron diffraction pattern collected at 298 K blown up to show where there would be peaks in a unit cell with an ordered honeycomb; the y axis is the measured intensity divided by the intensity of the largest peak. There is a peak around $Q = 1.5$ $Å^{-1}$ that indicates the presence of some in-plane ordering. The peak marked with an asterisk is thought to be from a small amount of $IrO_2$ impurity.

**Figure 4**: Neutron diffraction pattern of $Na_{2.8}ZnIr_2O_6$ collected at 300 K with fit to a monoclinic cell in space group C2/m. Observed intensity (red squares), calculated intensity (black line), difference calculation (blue line) and *hkl* reflections (green dashes) are shown.

**Figure 5:** Single crystal diffraction pattern of $Na_{1-x}Cu_{1/3}Ir_{2/3}O_2$, showing the *h0l* section of the reciprocal lattice. There is no streaking between the reflections, indicating a lack of stacking faults. The reciprocal lattice vectors are indicated.

**Figure 6**: Upper panel: The inverse magnetic susceptibility from 2 to 300 K of $Na_{0.62}Mn_{0.61}Ir_{0.39}O_2$ with $\chi_0$ subtraction (open squares) and without (closed squares). The data with the $\chi_0$ subtraction is fit to the Curie Weiss law. Inset: ZFC molar susceptibility of $Na_{0.62}Mn_{0.61}Ir_{0.39}O_2$ from 2 K to 300 K in a 1 T field. Center panel: The inverse magnetic susceptibility from 2 to 300 K of $Na_{0.80}Fe_{2/3}Ir_{1/3}O_2$ with $\chi_0$ subtraction (open squares) and without (closed squares). The data with the $\chi_0$ subtraction is fit to the Curie Weiss law. Inset: ZFC molar susceptibility of $Na_{0.80}Fe_{2/3}Ir_{1/3}O_2$ from 2 K to 300 K in a 0.1 T field. Lower panel: The inverse magnetic susceptibility from 2 to 300 K of $Na_{0.90}Co_{1/3}Ir_{2/3}O_2$ without $\chi_0$ subtraction (closed squares). The inverse susceptibility shows a change in slope at about 200 K; the linear regions above and below this temperature are fit to the Curie Weiss law. Inset: ZFC molar susceptibility of $Na_{0.90}Co_{1/3}Ir_{2/3}O_2$ from 2 K to 300 K in a 0.1 T field.

**Figure 7:** Upper panel: The inverse magnetic susceptibility from 2 to 300 K of $Na_{0.92}Ni_{1/3}Ir_{2/3}O_2$ with $\chi_0$ subtraction (open squares) and without (closed squares). The data with the $\chi_0$ subtraction is fit to the Curie Weiss law. Inset: ZFC molar susceptibility of $Na_{0.92}Ni_{1/3}Ir_{2/3}O_2$ from 2 K to 300 K in a 1 T field. Center panel: The inverse magnetic susceptibility from 2 to 300 K of $Na_{0.86}Cu_{1/3}Ir_{2/3}O_2$ with $\chi_0$ subtraction (open squares) and without (closed squares). The data with the $\chi_0$ subtraction is fit to the Curie Weiss law. Inset: ZFC molar susceptibility of

Na$_{0.86}$Cu$_{1/3}$Ir$_{2/3}$O$_2$ from 2 K to 300 K in a 1 T field. Lower panel: The inverse magnetic susceptibility from 2 to 300 K of Na$_{0.89}$Zn$_{1/3}$Ir$_{2/3}$O$_2$ with $\chi_0$ subtraction (open squares) and without (closed squares). The data with the $\chi_0$ subtraction is fit to the Curie Weiss law. Inset: ZFC molar susceptibility of Na$_{0.89}$Zn$_{1/3}$Ir$_{2/3}$O$_2$ from 2 K to 300 K in a 1 T field.

**Figure 8:** Frequency dependence of the AC magnetization measurements of a) Na$_{0.62}$Mn$_{0.61}$Ir$_{0.39}$O$_2$, b) Na$_{0.80}$Fe$_{2/3}$Ir$_{1/3}$O$_2$, c) Na$_{0.90}$Co$_{1/3}$Ir$_{2/3}$O$_2$, d) Na$_{0.92}$Ni$_{1/3}$Ir$_{2/3}$O$_2$, e) Na$_{0.86}$Cu$_{1/3}$Ir$_{2/3}$O$_2$ and f) Na$_{0.89}$Zn$_{1/3}$Ir$_{2/3}$O$_2$.

**Figure 9:** Comparison of the normalized susceptibility of the antiferromagnetic compounds Na$_{0.62}$Mn$_{0.61}$Ir$_{0.39}$O$_2$, Na$_{0.80}$Fe$_{2/3}$Ir$_{1/3}$O$_2$, Na$_{0.92}$Ni$_{1/3}$Ir$_{2/3}$O$_2$, Na$_{0.86}$Cu$_{1/3}$Ir$_{2/3}$O$_2$, and Na$_{0.89}$Zn$_{1/3}$Ir$_{2/3}$O$_2$. Negative deviations from the line representing Curie Weiss behavior indicate that ferromagnetic fluctuations increase as T$_f$ is approached from above. Na$_{0.80}$Fe$_{2/3}$Ir$_{1/3}$O$_2$ does not show negative deviations from Curie Weiss behavior.

**Figure 10:** The field dependent magnetization of Na$_{0.62}$Mn$_{0.61}$Ir$_{0.39}$O$_2$, Na$_{0.80}$Fe$_{2/3}$Ir$_{1/3}$O$_2$, Na$_{0.90}$Co$_{1/3}$Ir$_{2/3}$O$_2$, Na$_{0.92}$Ni$_{1/3}$Ir$_{2/3}$O$_2$, Na$_{0.86}$Cu$_{1/3}$Ir$_{2/3}$O$_2$, and Na$_{0.89}$Zn$_{1/3}$Ir$_{2/3}$O$_2$ measured at a temperature of 2 K.

**Figure 11:** Heat capacity of Na$_{0.92}$Ni$_{1/3}$Ir$_{2/3}$O$_2$ with nonmagnetic analog NaZn$_{1/3}$Pt$_{2/3}$O$_2$. The black squares correspond to the heat capacity of Na$_{0.92}$Ni$_{1/3}$Ir$_{2/3}$O$_2$ while the red circles correspond to the heat capacity of NaZn$_{1/3}$Pt$_{2/3}$O$_2$. The inset shows the difference in the heat capacity between the two compounds divided by temperature as a function of temperature.

**Figure 12:** The log of the resistivity vs. temperature for each compound.

**Figure 13:** The XAS spectra at the 3$d$ transition metal L$_{2,3}$ edges of all studied compounds together with corresponding reference compounds.

**Figure 14:** The Ir-L$_3$ edge of Na$_{0.86}$Cu$_{1/3}$Ir$_{2/3}$O$_2$ together with IrCl$_3$, Li$_2$IrO$_3$ and Sr$_2$ScIrO$_6$ as Ir$^{3+}$, Ir$^{4+}$ and Ir$^{5+}$ references.

**Figure 15**: a) $^{23}$Na NMR spectra of a powder sample of Na$_2$IrO$_3$ at selected temperatures. Sharp spikes indicated by arrows are field markers. Step-like structure visible 20 K is due to a powder averaging of the quadrupole satellites (see the text). b) $^{23}$Na NMR spectra of a powder sample of Na$_{0.92}$Ni$_{1/3}$Ir$_{2/3}$O$_2$ at different temperatures. Sharp spikes indicated by arrows are field markers. c) $^{23}$Na NMR spectra of a powder sample of Na$_{0.89}$Ni$_{1/3}$Ir$_{2/3}$O$_2$ at different temperatures. Sharp spikes indicated by arrows are field markers.

**Figure 16**: a) Temperature dependence of the width of the $^{23}$Na NMR spectra of Na$_2$IrO$_3$, Na$_{0.92}$Ni$_{1/3}$Ir$_{2/3}$O$_2$ and Na$_{0.89}$Zn$_{1/3}$Ir$_{2/3}$O$_2$ samples. Arrows indicate the ordering temperatures from the static magnetic data. Solid lines are guides for the eye. b) Temperature dependence of the $^{23}$Na longitudinal relaxation rate 1/T$_1$ of Na$_2$IrO$_3$ and Na$_{0.92}$Ni$_{1/3}$Ir$_{2/3}$O$_2$ samples. Arrows indicate the ordering temperatures from the static magnetic data. Solid lines are guides for the eye.

**Table 1**

| $Na_{1-x}M_{1/3}Ir_{2/3}O_2$ | M = Co | M=Co (minor) | M = Ni | M = Cu | M = Zn |
|---|---|---|---|---|---|
| a (Å) | 3.08181(3) | 3.0384(1) | 3.085747(8) | 3.094337(8) | 3.099036(1) |
| c (Å) | 16.0187(2) | 16.188(1) | 15.9959(2) | 16.02484(7) | 16.02140(1) |
| Na (3b) 0,0,1/2 | | | | | |
| Occ. | 0.904(1) | 0.570(4) | 0.9166(4) | 0.861(4) | 0.8935(6) |
| $B_{iso}$ | 2.48(4) | 4.2(2) | 1.28(3) | 1.90(3) | 1.709(4) |
| M/Ir (3a) 0,0,0 | | | | | |
| Occ. | 1 | 1 | 1 | 1 | 1 |
| $B_{iso}$ | 0.67(1) | 1.82(5) | 0.1417(6) | 0.189(6) | 0.361(1) |
| O (6c) 0,0,z | | | | | |
| z | 0.26726(2) | 0.2675(1) | 0.26755(3) | 0.26758(3) | 0.26688(1) |
| Occ. | 1 | 1 | 1 | 1 | 1 |
| $B_{iso}$ | 0.99(1) | 1.11(3) | 0.536(9) | 1.033(9) | 0.696(1) |
| $\chi^2$ | 1.457 | | 1.066 | 0.8827 | 0.8516 |
| $R_p$ | 38.8 | | 26.1 | 27.5 | 25.7 |
| $R_{wp}$ | 26.0 | | 21.4 | 21.2 | 19.6 |

**Table 2**

| $Na_{1-x}M_{2/3}Ir_{1/3}O_2$ | M = Mn | M=Fe | M = Fe (minor) |
|---|---|---|---|
| a (Å) | 3.011924(7) | 3.043442(2) | 3.02538(2) |
| c (Å) | 16.42513(3) | 16.27532(2) | 16.1951(2) |
| Na (3b) 0,0,1/2 | | | |
| Occ. | 0.617(1) | 0.797(1) | 0.93(2) |
| $B_{iso}$ | 2.02(1) | 0.999(9) | 0.73(9) |
| M/Ir (3a) 0,0,0 | | | |
| Occ. | 0.6119(2) / 0.3880(2) | 1 | 1 |
| $B_{iso}$ | 1.213(2) | 0.335(2) | 0.34(2) |
| O (6c) 0,0,z | | | |
| x | 0.04145(8) | 0 | 0 |
| y | 0.08289(8) | 0 | 0 |
| z | 0.270211(6) | 0.26813(2) | 0.26635(8) |
| Occ. | 1/3 | 1 | 1 |
| $B_{iso}$ | 0.940(9) | 0.283(6) | 0.14(2) |
| $\chi^2$ | 1.2691 | 1.483 | |
| $R_p$ | 48.2 | 16.2 | |
| $R_{wp}$ | 30.1 | 13.9 | |

**Table 3**

Na$_{2.8}$ZnIr$_2$O$_6$

| | | | |
|---|---|---|---|
| a (Å) | | | 5.3692(2) |
| b (Å) | | | 9.2964(3) |
| c (Å) | | | 5.6276(1) |
| β(°) | | | 108.358(2) |
| O1 | 8j:*x,y,z* | x | 0.767(1) |
| | | y | 0.164(1) |
| | | z | 0.7933(8) |
| | | Occ. | 1 |
| | | B$_{iso}$ | 0.48(9) |
| O2 | 4i:*x,0,z* | x | 0.725(2) |
| | | z | 0.182(2) |
| | | Occ. | 1 |
| | | B$_{iso}$ | 0.7(2) |
| Na2 | 4h:*1/2,y,1/2* | y | 0.320(3) |
| | | Occ. | 0.94(5) |
| | | B$_{iso}$ | 1.7(5) |
| Ir1 | 4g:*1/2,y,0* | y | 0.167(1) |
| | | Occ. | 0.92(2) |
| | | B$_{iso}$ | 0.36(3) |
| Zn2 | 4g:*1/2,y,0* | y | 0.167(1) |
| | | Occ. | 0.08(2) |
| | | B$_{iso}$ | 0.36(3) |
| Na1 | 2d:*1/2,0,1/2* | Occ. | 0.95(8) |
| | | B$_{iso}$ | 0.8(6) |
| Zn1 | 2a:*0,0,0* | Occ. | 0.85(4) |
| | | B$_{iso}$ | 0.36(3) |
| Ir2 | 2a:*0,0,0* | Occ. | 0.15(4) |
| | | B$_{iso}$ | 0.36(3) |
| $\chi^2$ | | | 0.9904 |
| R$_p$ | | | 26.3 |
| R$_{wp}$ | | | 20.7 |

**Table 4**

| Compound | $\chi_0$ | C | $\theta$ | $\mu_{eff}$ | $T_f$ | $\theta/T_f$ | $\Delta T_f/T_f\Delta\log\omega$ | Activation Energy | Oxidation State of 3d element |
|---|---|---|---|---|---|---|---|---|---|
| | (emu/Oe) | $\frac{\text{emu K}}{\text{mol f.u. Oe}}$ | (K) | ($\mu_B$/f.u.) | (K) | | | (eV) | |
| $Na_{0.62}Mn_{0.61}Ir_{0.39}O_2$ | 0.00007 | 0.99 | -94.0 | 2.81 | 14.6 | 6.4 | 0.089 | 0.58 | 3.73+ |
| $Na_{0.80}Fe_{2/3}Ir_{1/3}O_2$ | -0.0004 | 2.74 | -26.3 | 4.68 | 13.1 | 2.0 | 0.042 | 0.54 | 3+ |
| $Na_{0.90}Co_{1/3}Ir_{2/3}O_2$ | 0 | 0.36 | 21.5 | 1.70 | 10.6 | 3.2 | 0.074 | 0.41 | 3+ |
| | 0 | 0.57 | -81.8 | 2.14 | | | | | |
| $Na_{0.92}Ni_{1/3}Ir_{2/3}O_2$ | 0.00012 | 0.56 | -37.7 | 2.11 | 22.1 | 1.7 | 0.024 | 0.63 | 2+ |
| $Na_{0.86}Cu_{1/3}Ir_{2/3}O_2$ | 0.00005 | 0.46 | -47.1 | 1.91 | 5.5 | 8.6 | 0.056 | 0.32 | 2+ |
| $Na_{0.89}Zn_{1/3}Ir_{2/3}O_2$ | -0.00015 | 0.24 | -27.0 | 1.39 | 2.7 | 10 | 0.091 | 0.76 | (2+) |

Figure 1

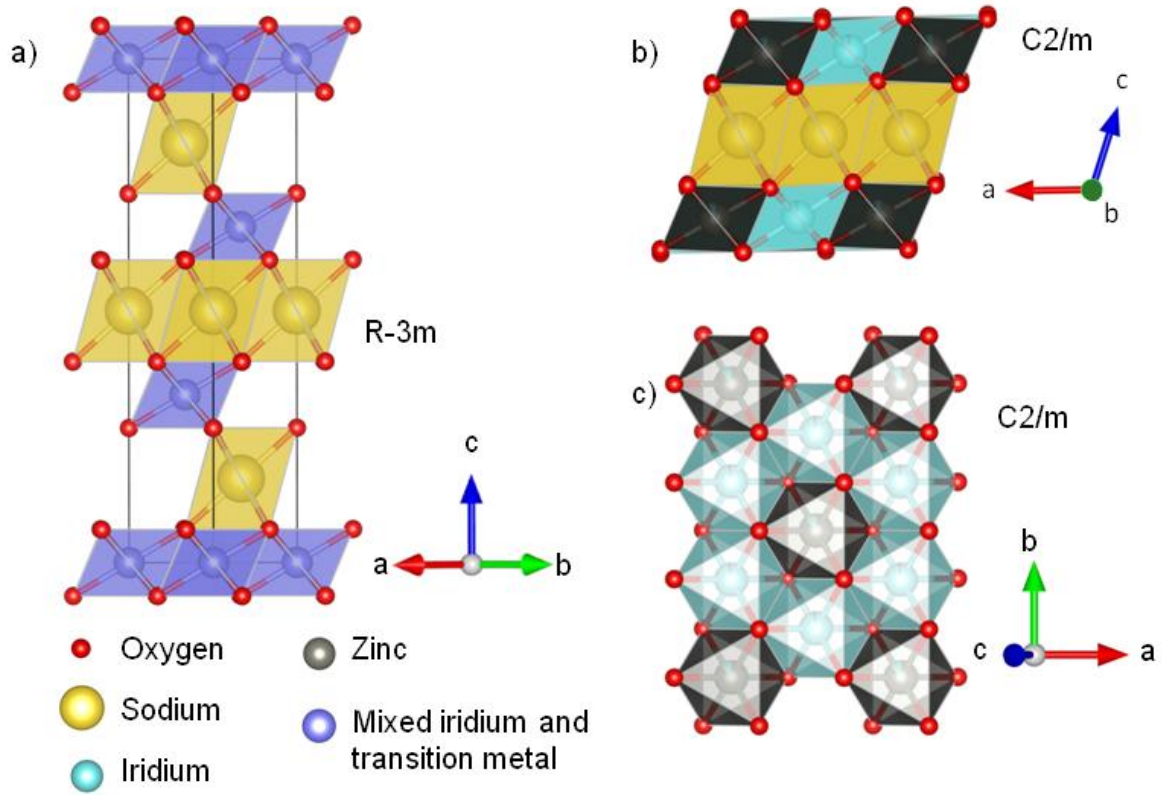

Figure 2

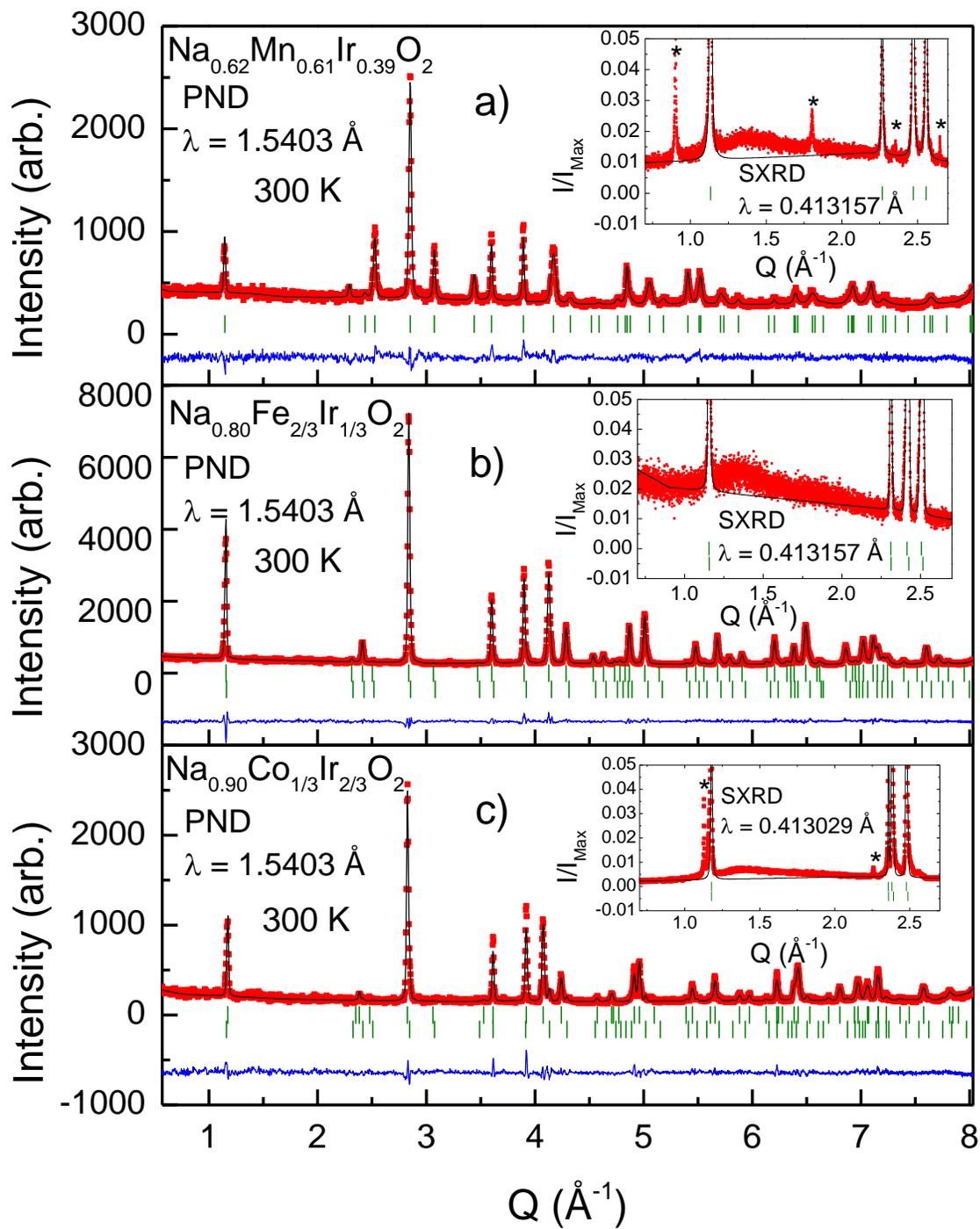

Figure 3

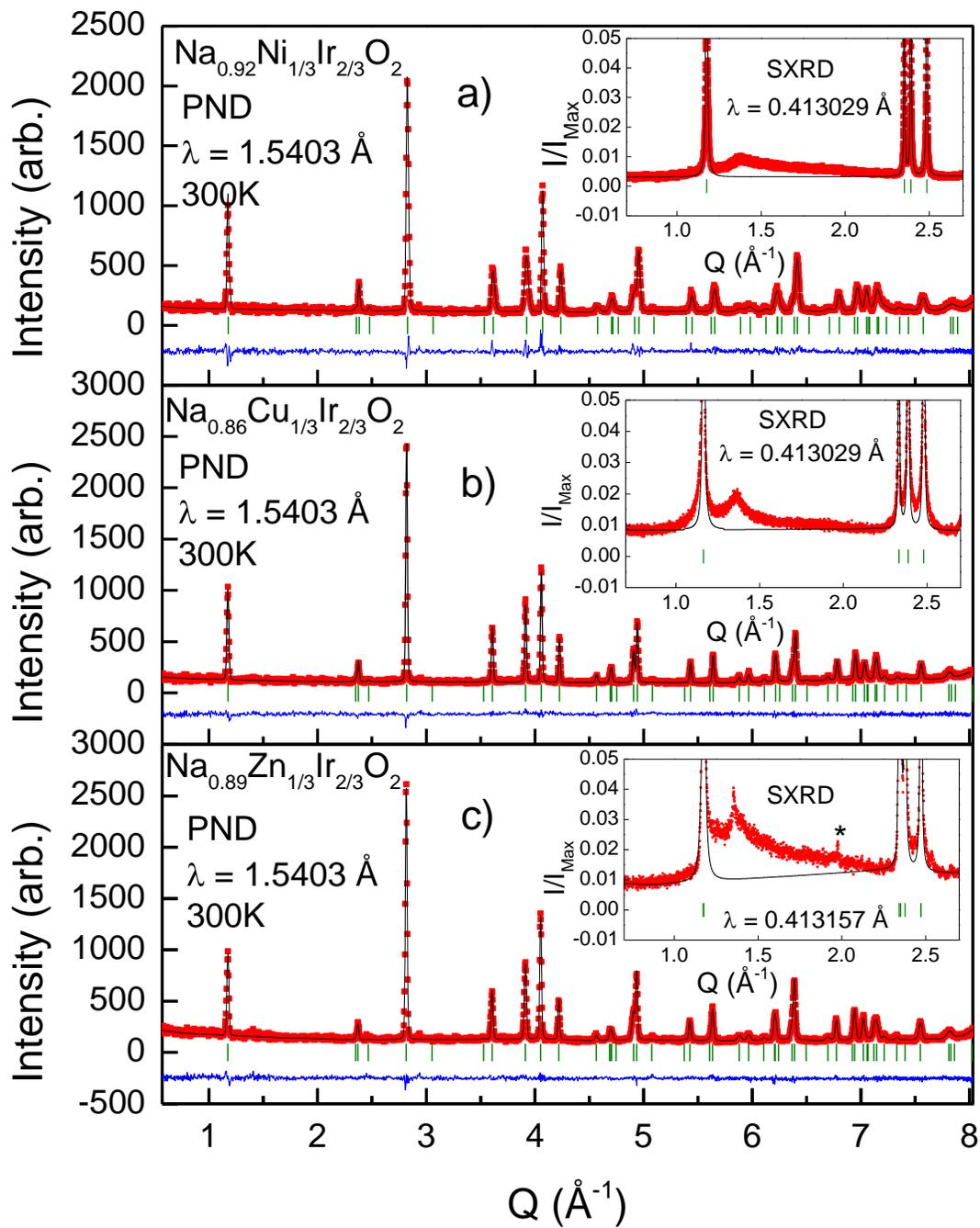

Figure 4

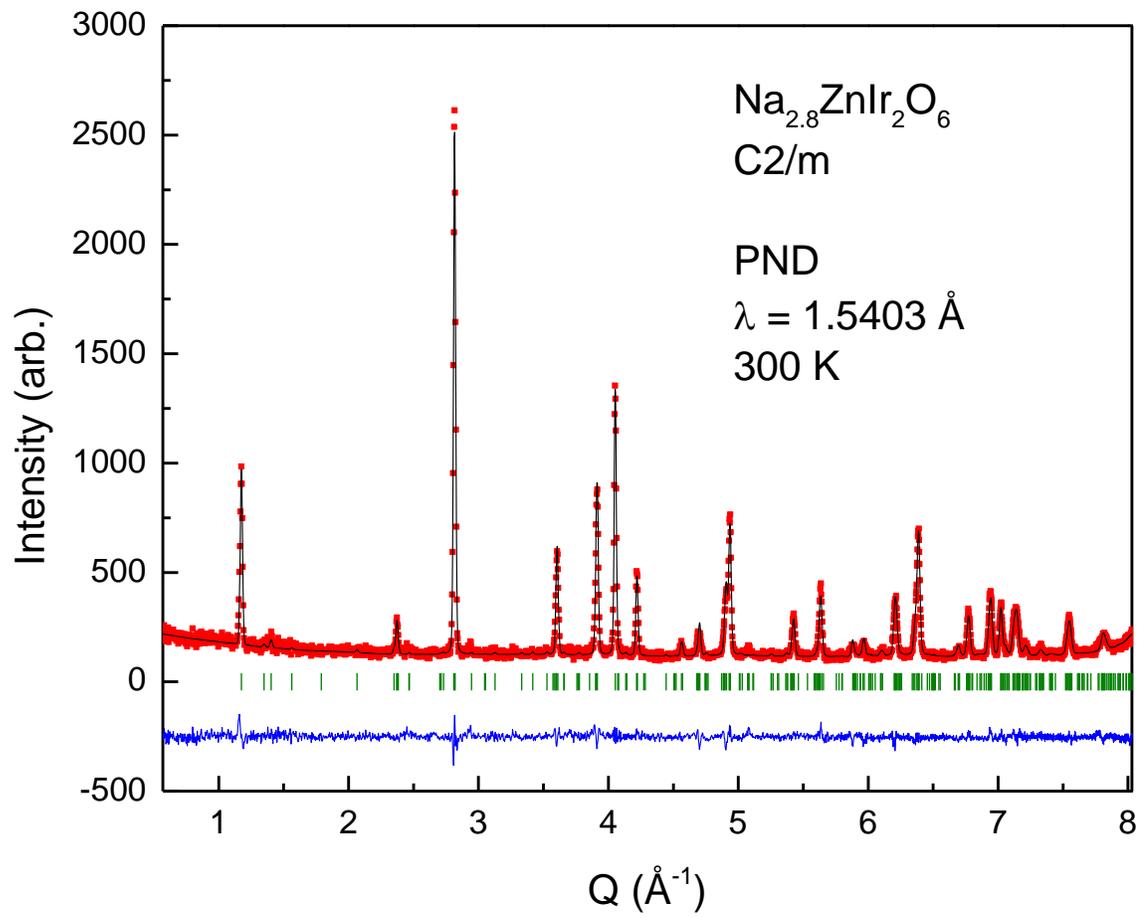

Figure 5

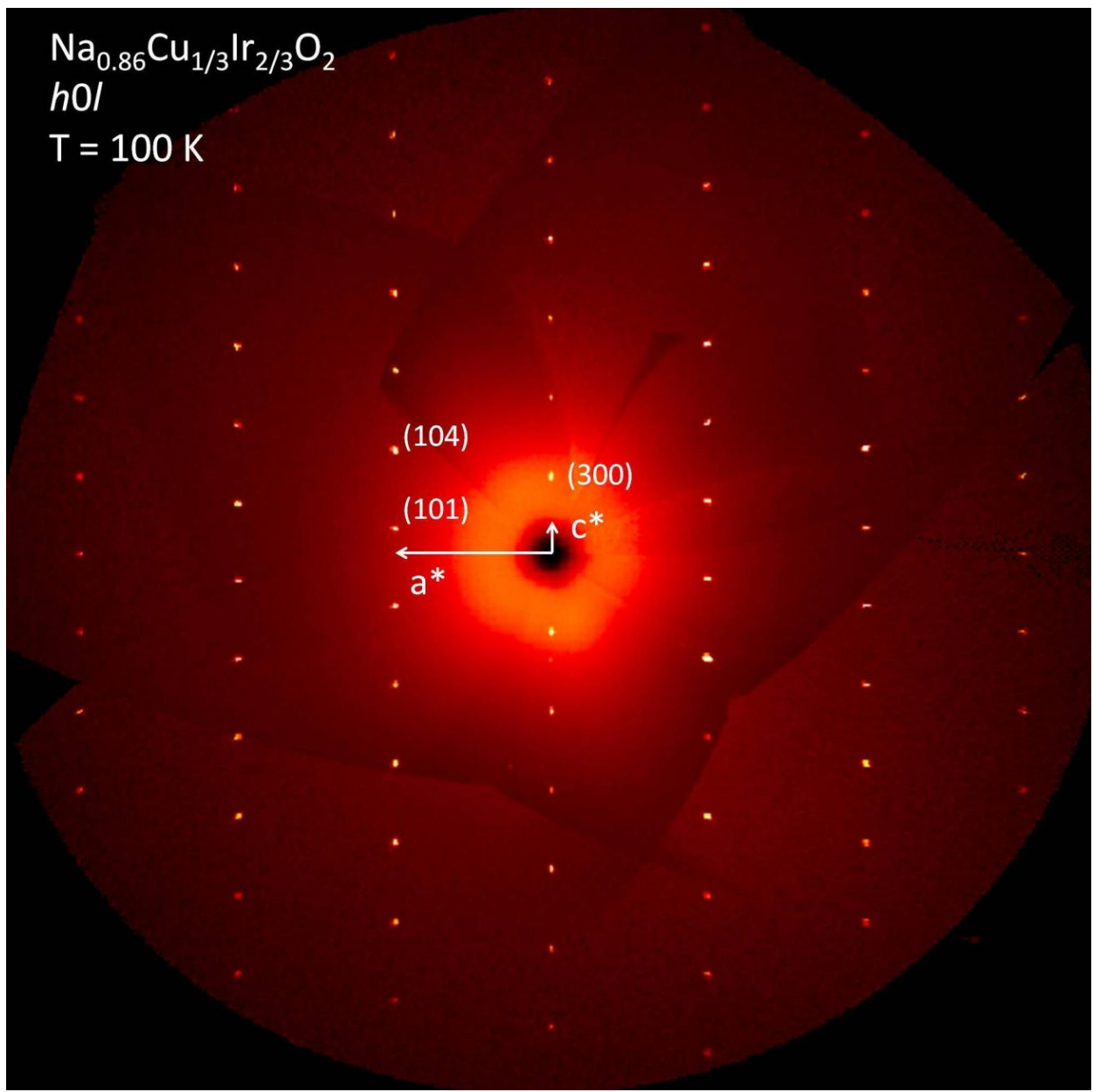

Figure 6

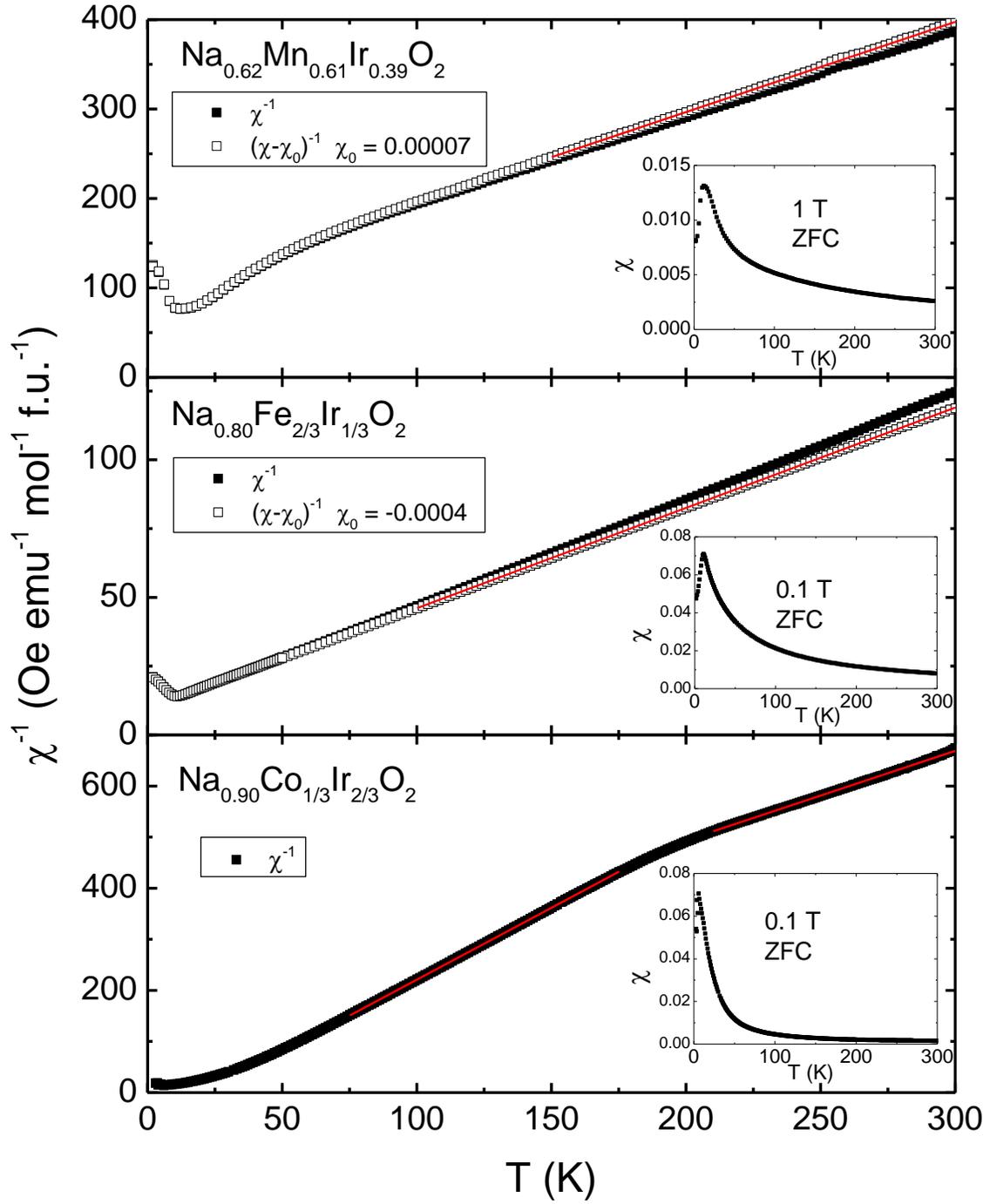

Figure 7

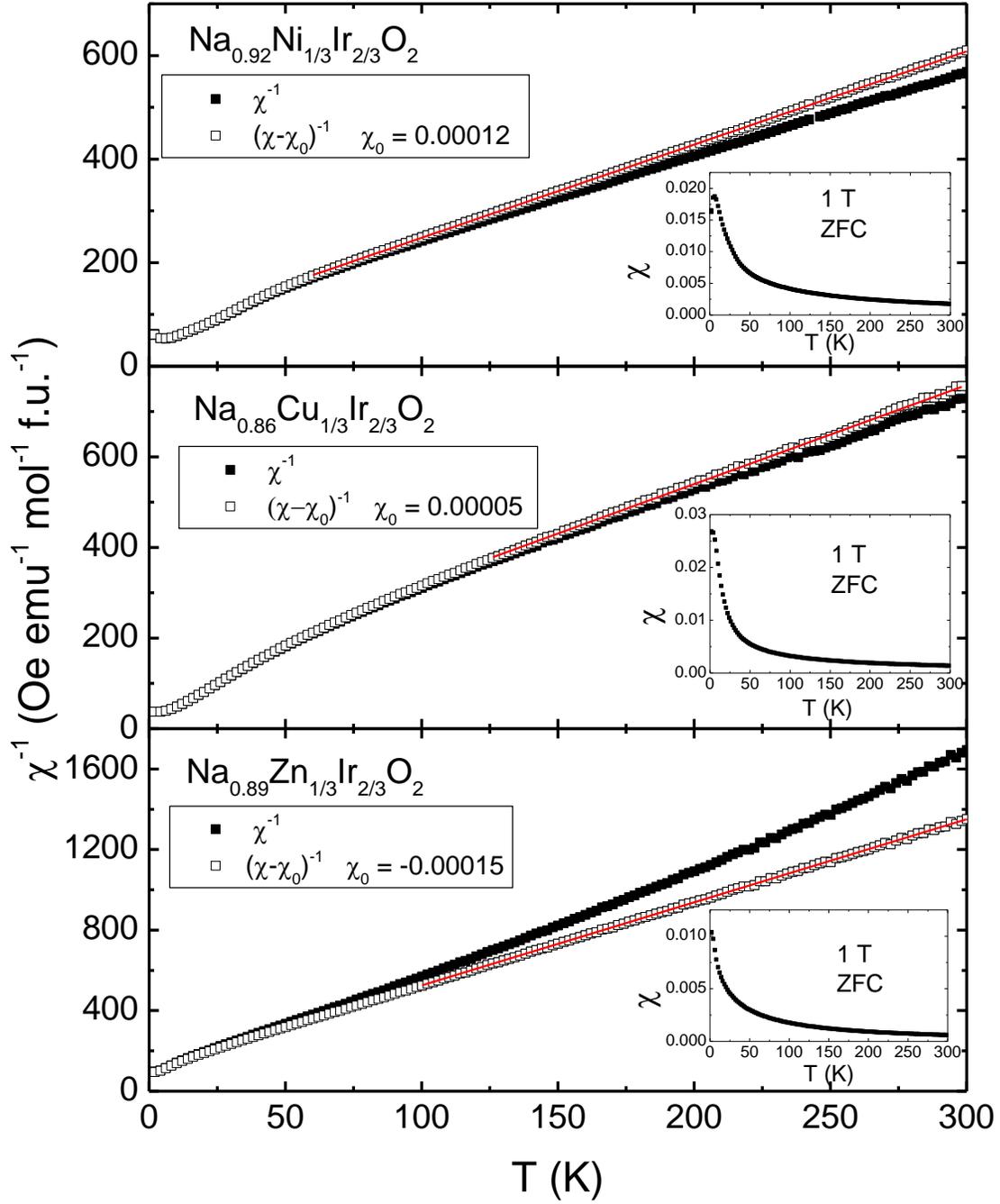

Figure 8

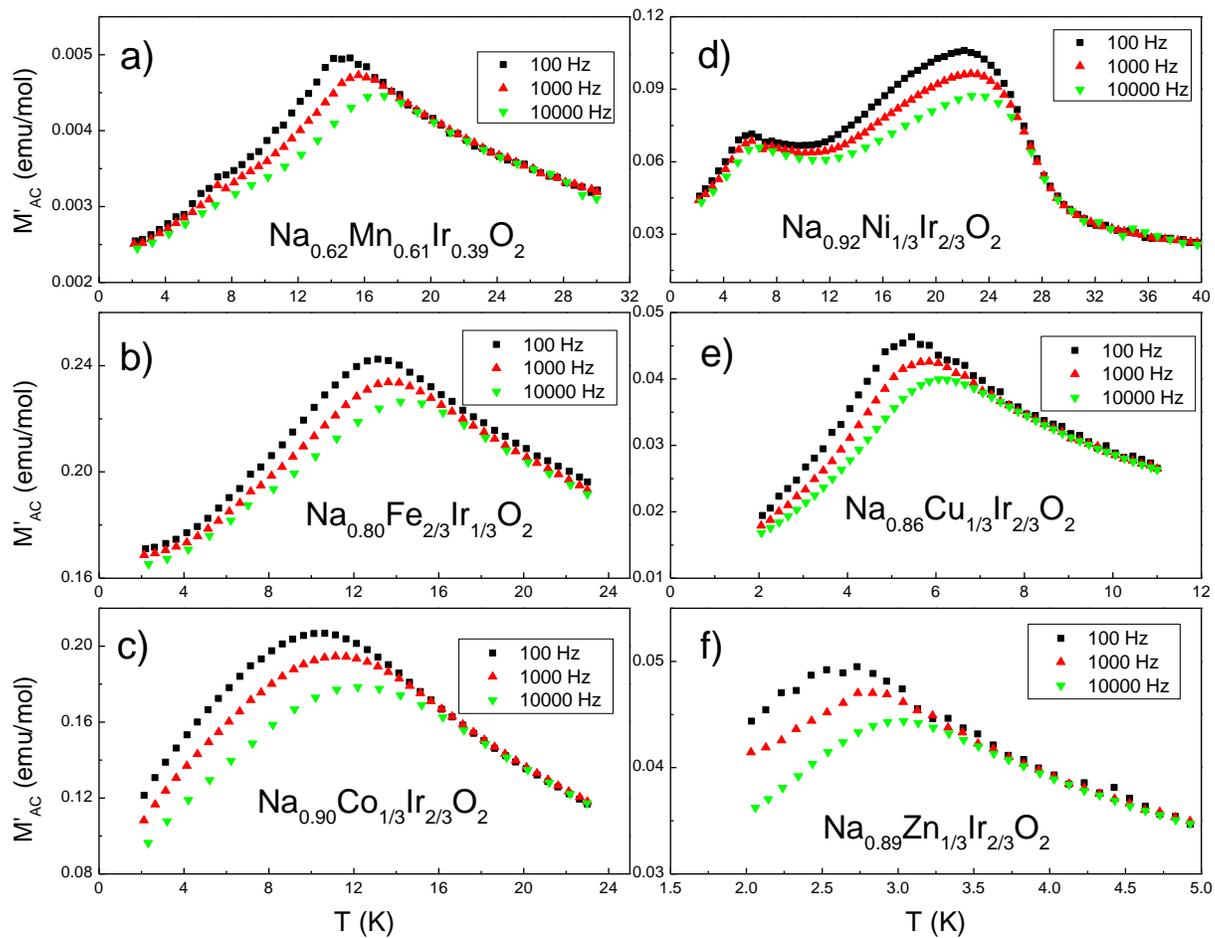

Figure 9

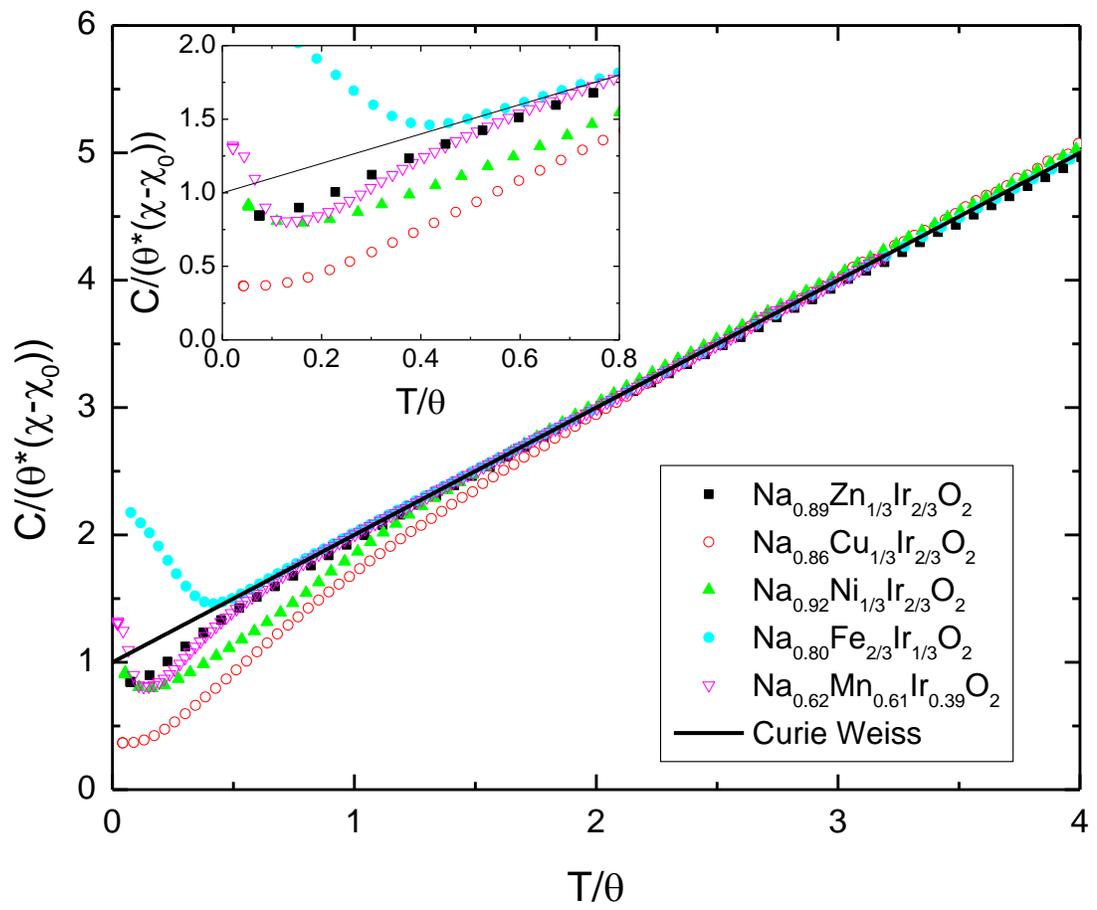

Figure 10

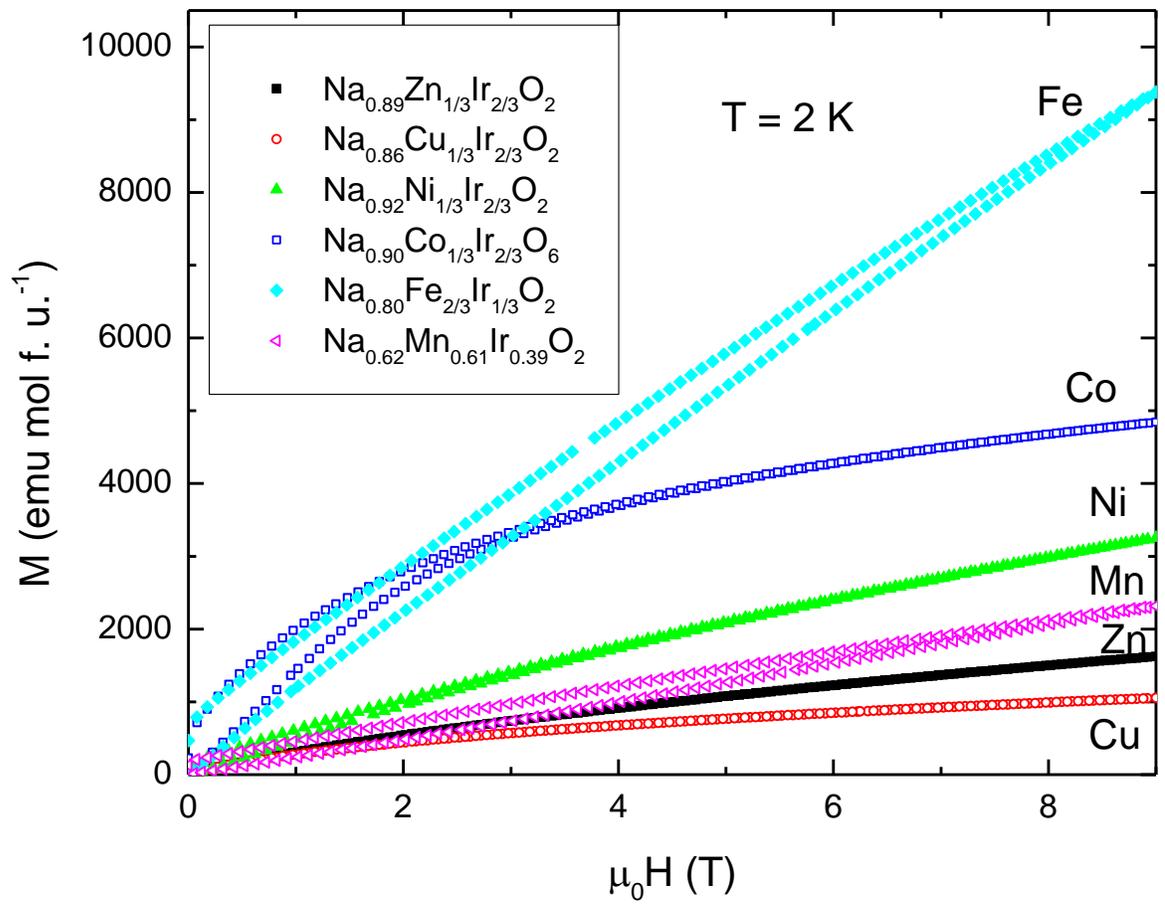

Figure 11

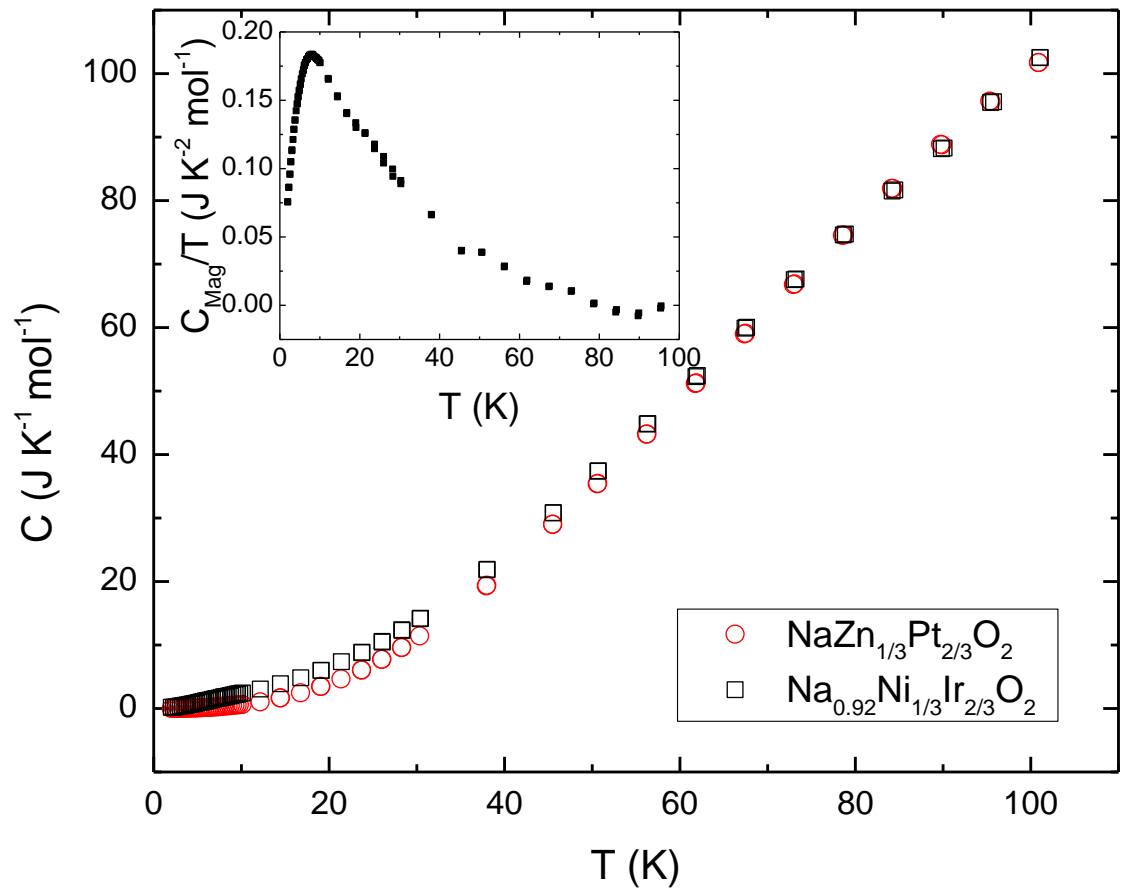

Figure 12

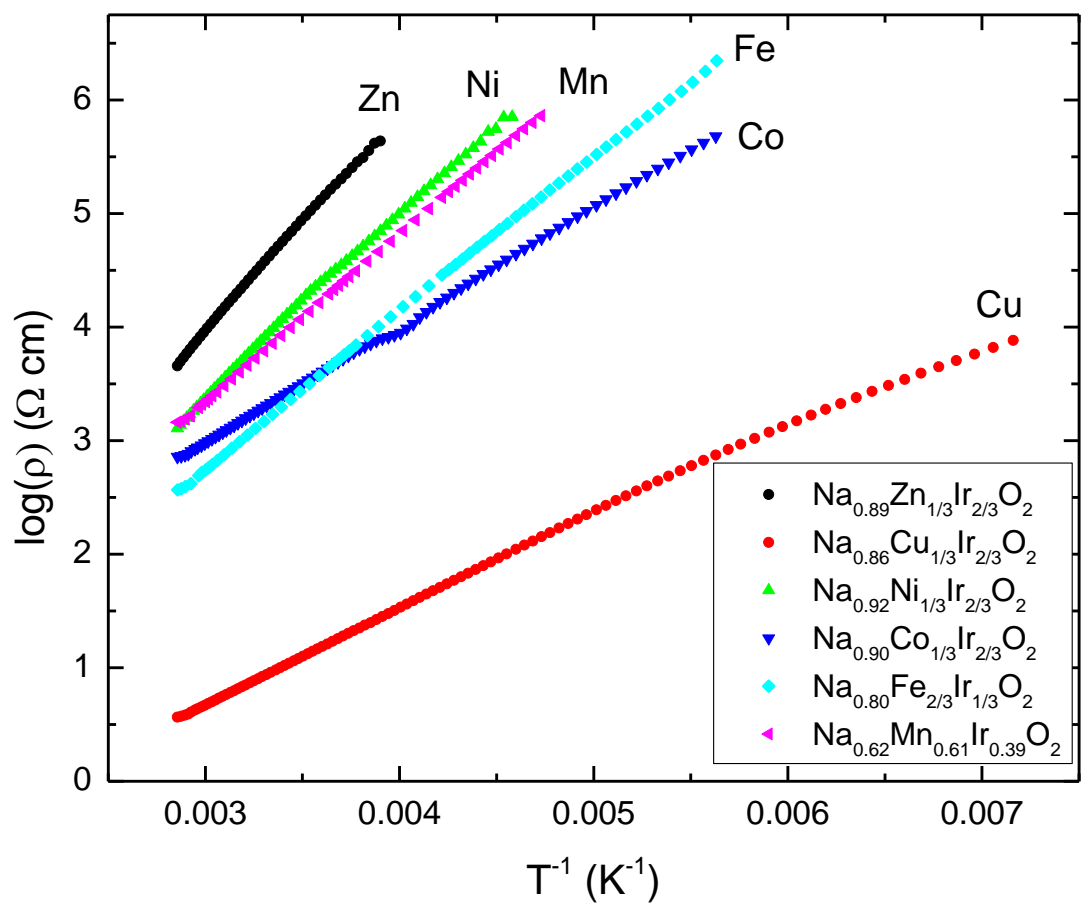

Figure 13

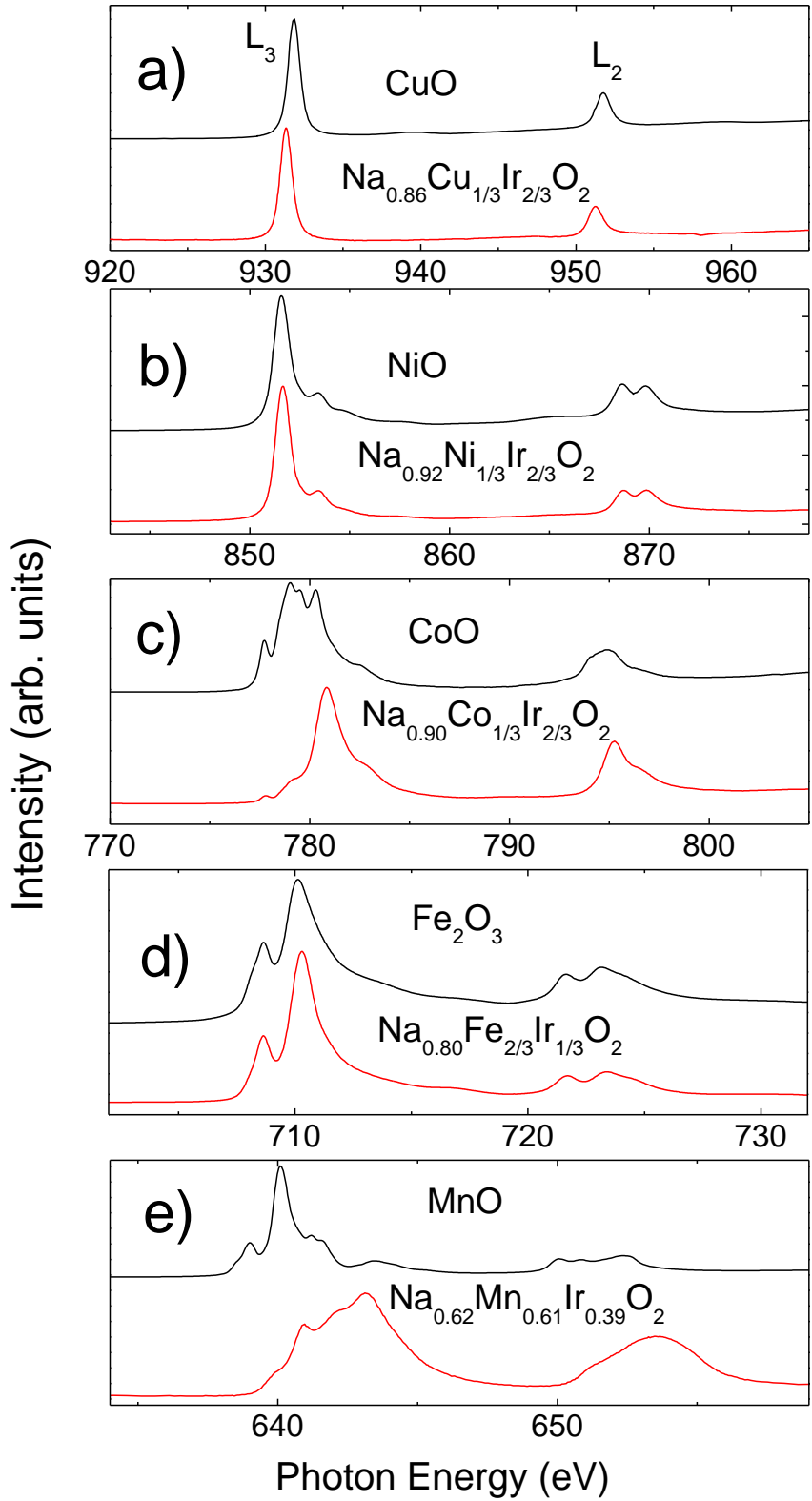

Figure 14

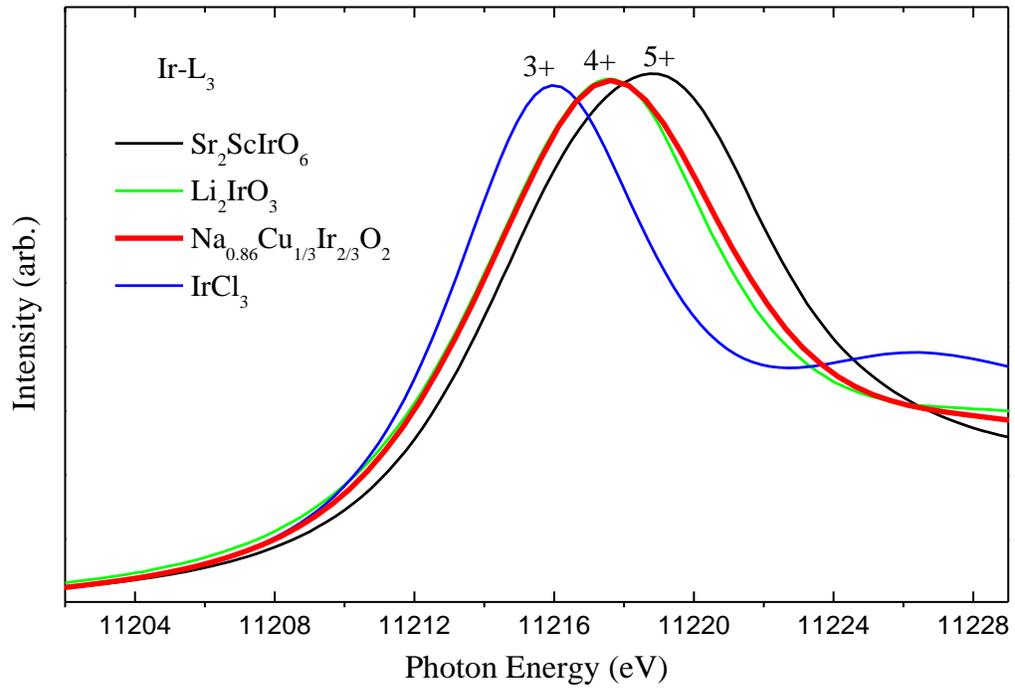

Figure 15

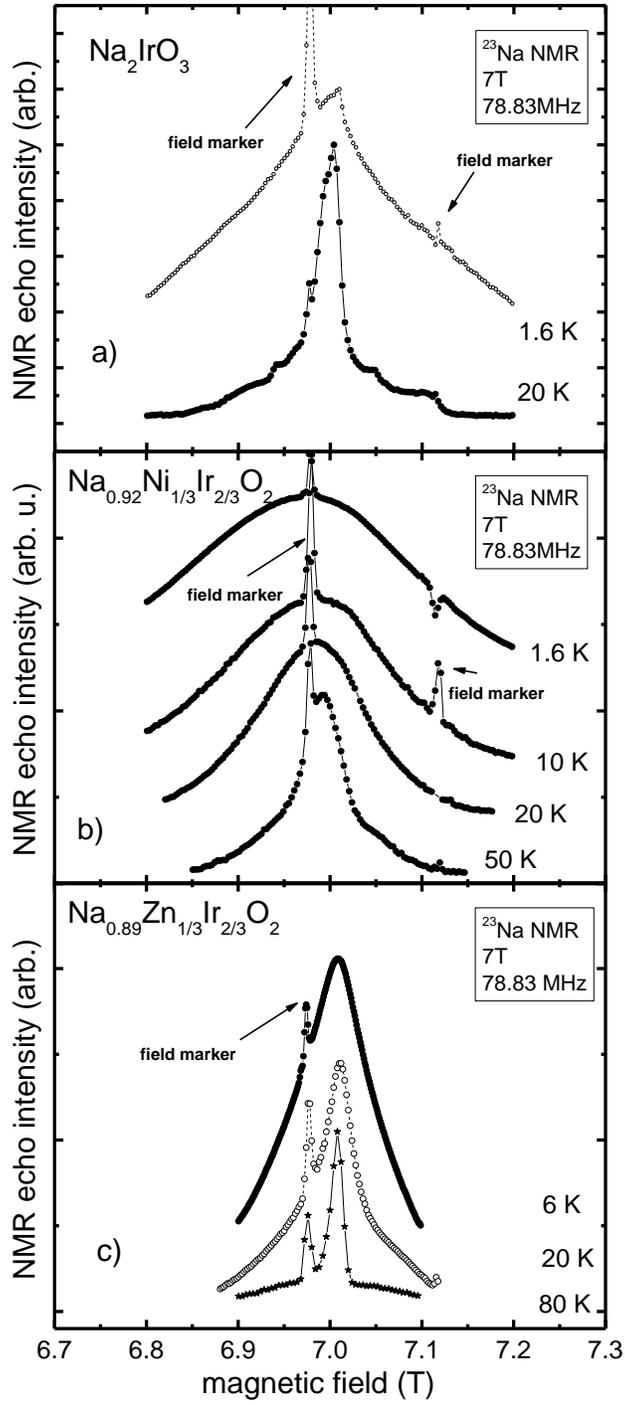

Figure 16

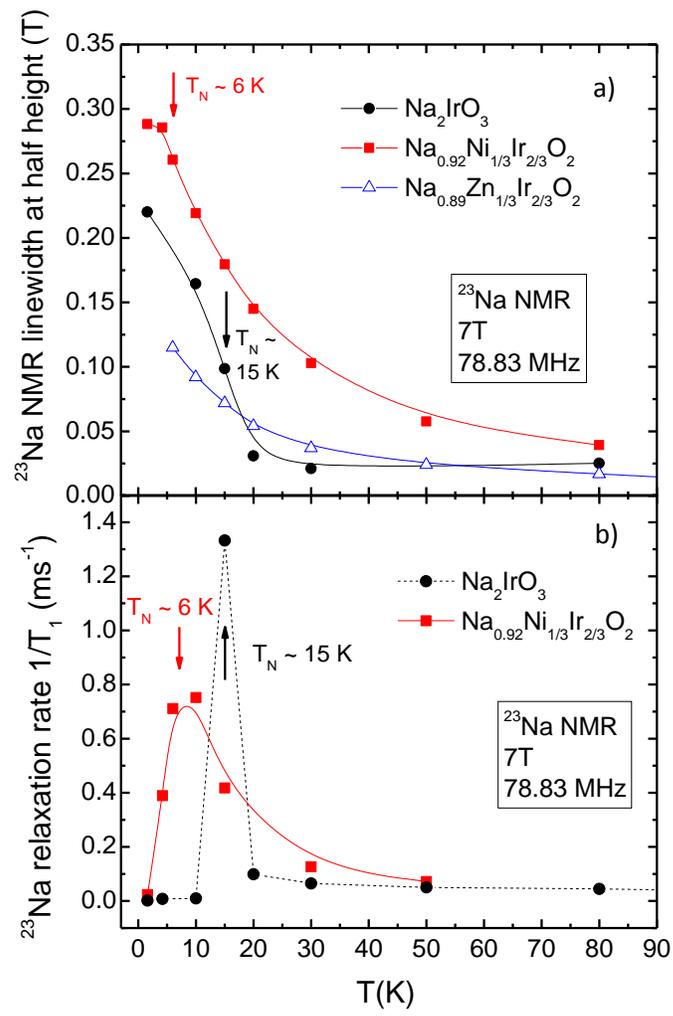